\documentclass[fleqn,10pt]{wlscirep}
\usepackage[T1]{fontenc}  
\usepackage{babel} \usepackage{txfonts}
\usepackage{epsfig,epsf,psfrag} 
\usepackage{graphicx} 
\usepackage{subfigure}
\usepackage{pslatex}
\usepackage{epic,eepic} 
\usepackage{color,pstcol}
\usepackage{pstricks} 
\usepackage{times}
\renewcommand\thesection{\arabic{section}}
\renewcommand{\thesubsection}{\thesection.\arabic{subsection}}
 \vspace{-10.0cm}  
  \title{Designing a highly efficient graphene quantum spin heat engine} 
  \author[1]{Arjun Mani}
  \author[1]{Subhajit Pal}
 \author[1,*]{Colin Benjamin} \affil[*]{colin.nano@gmail.com}\affil[1]{School of Physical Sciences, National Institute of Science Education \& Research, HBNI, Jatni-752050,\ India }
\begin{abstract}
 We design a quantum spin heat engine using  spin polarized ballistic modes generated in a strained graphene monolayer doped with a magnetic impurity. We observe remarkably large efficiency and large thermoelectric figure of merit both for the charge as well as spin variants of the quantum heat engine. This suggests the use of this device as a highly efficient quantum heat engine for charge as well as spin based transport. Further, a comparison is drawn between the device characteristics of a graphene spin heat engine  against a quantum spin Hall heat engine. The reason being edge modes because of their origin should give much better performance. In this respect we observe our graphene based spin heat engine can almost match the performance characteristics of a quantum spin Hall heat engine. Finally, we show that a pure spin current can be transported in our device in absence of any charge current.
 \end{abstract}
\begin{document}
\flushbottom
\maketitle
\section{Introduction}
The efficacy of quantum heat engines(QHE) at the nanoscale has been made more than obvious in the past half decade \cite{bauer-rev}. From being useful in schemes for removal of excess heat in nanosystems to novel nano heat engines which produce huge amounts of power they have been one of the most productive areas of research \cite{Bauer}. {Graphene as a thermoelectric material has a very small thermoelectric figure of merit $ZT$ around 0.1-0.01, which is much smaller than the most efficient thermoelectric material $Bi_2Te_3$, see Refs.~\cite{kim, yuri}. This is due to its large thermal conductance and absence of any band gap. In some recent works, a moderate improvement of the thermoelectric figure of merit $ZT$ is noticed in graphene based systems. This improved thermoelectric figure of merit $ZT$ of around 2.5-3 has been observed in 2D graphene systems with disorder \cite{anno,tran} or isotopes\cite{tran} or nanopores\cite{sadeghi} or by nano-patterning the graphene surface\cite{kim}. This thermoelectric figure of merit observed in 2D graphene system is still smaller than that of the heat engine based on spin wave ferromagnetic system, see Ref.~\cite{Bauer}}. {In one of our previous works, see Ref.~\cite{arjun}, we designed a QHE based on monolayer graphene system to improve the overall performance and discussed the effect of strain. In Ref.~\cite{arjun}, we have only discussed the charge thermoelectric properties of strained graphene, the spin thermoelectric properties are not examined. Here, we mainly concentrate on the interplay between the roles of strain and spin flip scattering due to the magnetic impurity. In this work we prescribe a recipe to design a quantum spin heat engine(QSHE) using spin polarized ballistic modes in strained graphene. We find giant thermoelectric factors of around 50 for both charge as well as spin based transport. 

There have been a few papers on marrying spin transport into heat engines, mention may be made of Ref.~\cite{Bauer} wherein both the spin as well as charge thermoelectric factors are calculated along with the power and efficiency of both charge as well as spin heat engines. {In Ref.~{\cite{shen}}, charge/spin thermoelectric properties of a carbon atomic chain sandwiched between two ferromagnetic zigzag graphene nanoribbon is studied at various range of temperatures (from $0-400$ K).} In Ref.~\cite{ramshetti}  the spin and charge thermoelectric figure of merits for a ferromagnetic graphene based QHE is calculated. Finally, in Ref.~\cite{pohao} the authors calculate the thermoelectric figure of merit as well as power output in  a graphene based heat engine with spin polarized edge modes. However, what is unique to our work is that the same graphene based heat engine under strain and doped with a magnetic impurity can work as a highly efficient charge as well as spin heat engine. We also show that our device generates a charge power almost twice than what is seen in it's closest competitor, see Ref.~\cite{pohao}. {In some of the recent works, see Refs.~ \cite{zeng,xiao, yun}, the possibility of graphene to work as spin caloritronic devices is also studied, where graphene nanoribbon devices are engineered to generate large spin currents on application of a temperature difference at the two opposite edges of the system. In our device too, it can be shown that pure spin currents can be generated on application of only temperature difference.} In a previous work we had dealt in detail with a quantum spin Hall based QSHE\cite{hel}. { The charge and spin thermoelectric properties of graphene nanoribbon has also been studied in a recent paper\cite{ing} in the ballistic transport regime similar to us, but in their model they have spin-orbit coupling of Rashba type instead of a magnetic impurity. In Ref.~\cite{ing}, two types of spin currents are studied, one along the direction of applied thermal bias and the other one in the transverse direction, while in this work we studied the spin currents only in the direction of the thermal bias.} }{ In Refs.~\cite{Zhen,Zhan}, it has been shown that in presence of strong spin orbit coupling and exchange interaction due to the presence of magnetic impurities, quantum anomalous Hall effect can be observed in graphene. Since intrinsic spin orbit interaction of graphene is very small, in Refs.~\cite{Zhen,Zhan} it has been introduced via a substrate and magnetic impurities. In Ref.~\cite{Zhen}, it has been shown that Rashba spin splitting of magnitude $225$ meV can be observed if Nickel is used as substrate. Thus for the experimental realization of our model we can utilize a different substrate with low spin orbit interaction, such as Silicon or Germanium. We also choose the magnetic impurity such that spin orbit interaction introduced by it is minimum. In Ref.~\cite{Zhan}, it has been shown that when magnetic impurities are distributed all over the sample then a strong Rashba spin splitting results. In our case, on the other hand we only have a single magnetic impurity located at $x=0$. Thus, we can safely neglect the spin orbit interaction in our system, see Fig.~1. In some of the recent works, see Refs.~\cite{yan, miao, yuro}, the thermoelectric power of a graphene nanoribbon has also been studied in presence of a strong magnetic field. In these papers, along with the Seebeck coefficient, they have also studied the Nernst coefficient, which is the measure of charge current generated along the transverse direction.} {Further, in Refs.~\cite{pena,mun}, the graphene monolayer system and a single Dirac particle trapped in a infinite potential well are used to design the quantum analogue of classical Otto engine and classical Carnot engine respectively. While these quantum heat engines work as the cyclic heat engines, our model consists of strained graphene layer and works as a steady state quantum heat engine. The working principle of cyclic heat engine is that after a 
complete cycle all the parameters return back to their initial position via a reversible process, and since a reversible process take infinite time to complete the cycle, the output power generated in these heat engines are almost zero. For steady state heat engines the output power is finite and is generated via steady state flows of microscopic particles.}

In this work, we also compare the performance of charge/spin quantum heat engine based on monolayer graphene with that based on edge modes in quantum spin Hall systems. The reason behind this comparison is that while a graphene based QSHE relies on ballistic modes, a quantum spin Hall heat engine will rely on edge modes. Edge modes are observed at the edges of topological insulators (quantum spin Hall system), while the best place to see ballistic modes is monolayer graphene. Edge modes are different from ballistic modes in that the transmission probability for edge mode transport is unity, i.e., there is no back reflection of the edge modes when they encounter any impurity present within the sample, while for ballistic mode transport, transmission probability can be less than unity, i.e., ballistic modes are not completely immune to backscattering due to impurity present within the sample. This comparison will give a better perspective on application of edge modes verses ballistic modes in thermoelectrics.

The reason we aim to design a quantum spin heat engine(QSHE) in graphene is that the prospect for device realization is high. Since in graphene, electronic transport can be very easily tuned by a gate voltage alone. In our model graphene QSHE too, by optimizing the parameters, heat can be converted to a spin polarized charge current as well as a pure spin current similar to Ref.~\cite{sothmann2}. { A bandgap or a conduction gap present in a sample can enhance the Seebeck coefficient and thus the performance of the quantum heat engine. It can be explained in this way. In presence of a conduction gap or a band gap the electrical conductance reduces but the Seebeck coefficient increases, since Seebeck coefficient is inversely proportional to the electrical conductance (see Eq.~(2) of the main manuscript). However, there are some restrictions. To get a finite Seebeck current one has to break both the left-right symmetry, i.e., $T_{12}V_2\neq T_{21}V_1$ although time reversal symmetry ($T_{12}=T_{21}$) is not broken, and electron-hole symmetry. When one of the contacts of a particular system is hotter than all other contacts in that system then a electron-hole pair created at that contact which traverses to other contacts carries the excess heat energy of that contact. Now, since both electron and holes in a pair are transmitting to the other contacts, one will not get any current. To get a finite current one has to break the electron-hole symmetry, which is only possible when the transmission function is energy dependent. This energy dependent transmission function will act as a rectifier, which creates an asymmetry between electron and hole transport. Thus, to get a better thermoelectric (TE) performance one has to have a conduction gap and specific energy dependent transmission function as well. In a related paper, see Ref.~\cite{kao}, it has been concluded that a delta like transmission function in energy helps in reaching the Carnot limit in those quantum heat engines. Since $MoS_2$ and black phosphorus both have a band gap, and if they too satisfy the delta like transmission function in energy, then quantum heat engines based on these materials will also have a large Seebeck coefficient as well as large efficiency.}

 The manuscript is arranged as follows in section 2 we delve into the theory needed to understand the quantum spin heat engine. Next in section 3 we introduce our model which consists of a strained mono-layer of graphene embedded with a magnetic impurity. In section 4 we discuss the results of our work with a few plots of the spin and charge Seebeck coefficients, of the charge/spin thermoelectric factors and  of the efficiency and power of our model spin heat engine, we also deal with the a novel application of our spin heat engine to generate pure spin current and we delve deep into the causes behind the novel effects seen by plotting the bandstructure in presence of spin flip scattering and strain. Section 5 deals with the experimental realization of our proposed device. We conclude with an experimental realization of our proposed device in section 6 along with a table which compares the spin/charge efficiency, thermoelectric factors of our device with edge mode based quantum spin Hall heat engine.
\section{Theory of the quantum spin heat engine}
\subsection{Spin Seebeck coefficient}
 The aim of our work as stated in the introduction is to design a quantum spin  heat engine using a strained graphene layer embedded with a magnetic impurity. It goes without saying that our device acts as a quantum charge heat engine too. For this we begin by defining the thermoelectric properties of our graphene system in the linear transport regime- the electric and heat currents are linearly proportional to the applied biases be it electric or thermal. As is well known electrons in graphene can be both valley ($K/K'$) polarized as well as spin ($\uparrow/\downarrow$) polarized\cite{firoz,firoz2}. The linear dependencies can be expressed as follows \cite{ramshetti, benetti, sothmann}-
\begin{equation}\label{current}
\left(\begin{array}{c} j_s^v \\ j_s^{q, v}\end{array}\right)=\left(\begin{array}{cc} L_s^{11, v} & L_s^{12, v}\\ L_s^{21, v} & L_s^{22, v} \end{array}\right) \left(\begin{array}{c} -\mathcal{E}\\ -\Delta T \end{array}\right),
\end{equation}
where $j_s^v$ and $j_s^{q, v}$ are the electric and heat currents for spin $'s'$ electrons ($s\in{\uparrow, \downarrow}$) respectively, and $v$ is for $K/K'$ valley, $L_{ij}$ with $i,j \in 1,2$ represents the Onsager coefficients for a two terminal thermo-electric system. 
The electric response due to a finite temperature difference $\Delta T$ across the graphene layer is denoted as the Seebeck coefficient while the heat current generated due to the applied bias voltage $\mathcal{E}$ across graphene layer is denoted as Peltier coefficient. Using Eq.~(\ref{current}) these aforesaid coefficients can be expressed as-
\begin{equation}\label{Seebeck1}
S_s^v=-\frac{L_s^{12, v}}{L_s^{11, v}}, \quad\text{and}\qquad P_s^v=\frac{L_s^{21, v}}{L_s^{11, v}}.
\end{equation}

Due to the additional spin($s$) and valley($\nu$) degrees of freedom for electrons in graphene the charge($S_{ch}^\nu$) and spin Seebeck($S_{sp}^\nu$) and Peltier coefficients($P_{ch}^{\nu}, P_{sp}^{\nu}$) for any valley ($\nu=K/K'$) can be written as\cite{Bauer}- 
\begin{eqnarray}
S_{ch}^v&=&\frac{G_{\uparrow}^vS_{\uparrow}^v+G_{\downarrow}^vS_{\downarrow}^v}{G_{\uparrow}^v+G_{\downarrow}^v} \text{ and }  S_{sp}^v=\frac{G_{\uparrow}^vS_{\uparrow}^v-G_{\downarrow}^vS_{\downarrow}^v}{G_{\uparrow}^v+G_{\downarrow}^v},\\
P_{ch}^v&=&\frac{G_{\uparrow}^vP_{\uparrow}^v+G_{\downarrow}^vP_{\downarrow}^v}{G_{\uparrow}^v+G_{\downarrow}^v} \text{ and }  P_{sp}^v=\frac{G_{\uparrow}^vP_{\uparrow}^v-G_{\downarrow}^vP_{\downarrow}^v}{G_{\uparrow}^v+G_{\downarrow}^v}.
\end{eqnarray}
The sum over both valleys ($K$ and $K'$) gives the total charge/spin Seebeck and Peltier co-efficients-
\begin{eqnarray}
S_{ch}=S_{ch}^K+S_{ch}^{K'} \text{ and }S_{sp}=S_{sp}^K+S_{sp}^{K'},\\
P_{ch}=P_{ch}^K+P_{ch}^{K'} \text{ and }P_{sp}=P_{sp}^K+P_{sp}^{K'}.
\end{eqnarray}
To simplify matters, the Onsager co-efficient matrix in Eq.~(\ref{current}), relating electric and heat currents to temperature difference and applied electric bias, can be rewritten as follows \cite{ramshetti,shefranuk,dolphas}-
\begin{equation}\label{onsager}
\left(\begin{array}{cc} L_s^{11, v} & L_s^{12, v}\\ L_s^{21, v} & L_s^{22, v} \end{array}\right) =\left(\begin{array}{cc} \mathcal{L}_s^{0, v} & \mathcal{L}_s^{1, v}/eT\\ \mathcal{L}_s^{1, v}/e & \mathcal{L}_s^{2, v}/e^2T \end{array}\right)
\end{equation}
with,
\begin{eqnarray}\label{con}
\mathcal{L}_s^{\alpha, v}=G_0\int_{-\pi/2}^{\pi/2}\!\!\!d\phi \cos\phi \int_{-\infty}^{\infty}\!\!\!d\epsilon(-\frac{\partial f}{\partial \epsilon})\frac{|\epsilon|}{\hbar v_f}(\epsilon-\mu)^\alpha \mathcal{T}_s^v(\epsilon,\phi)\nonumber\\
\end{eqnarray}
herein $G_0=(e^2/\hbar)(W/\pi^2)$, with $W$ being the width of graphene layer in $y-$ direction, $\mathcal{L}_{s}^{0, v}=G_{s}^v$ is conductance of graphene electrons with spin $s$, in valley $v$ \cite{dolphas}. $\phi$ is the angle at which the electron is incident, $\epsilon$ is the energy of the electron, $f$ is the Fermi-Dirac distribution, $\mu$ is the Fermi energy and $\mathcal{T}_s^{\nu}(\epsilon, \phi)$ is the transmission probability for spin $s$ electrons through strained graphene for valley $\nu$. To calculate the Onsager coefficients $L_{s}^{ij,\nu}$ in Eq.~(\ref{current}),  one first has to calculate the transmission probability $\mathcal{T}_s^{\nu}(\epsilon,\phi)$ and then after calculating the Onsager coefficients $L_s^{ij,v}$ in Eq.~(\ref{current}), we calculate efficiency and power of our quantum spin heat engine. To do that we need to write the response matrix in terms of electric charge($J_{ch}$) and spin($J_{sp}$) currents as well as heat current ($J_Q$), which can be calculated from Eq.~(\ref{current}) by using the relations- $J_{ch}=j_{\uparrow}+j_{\downarrow}$, $J_{sp}=j_{\uparrow}-j_{\downarrow}$ and $J_{Q}=j_{\uparrow}^q+j_{\downarrow}^q$  as follows \cite{Bauer}- 
\begin{equation}\label{aa}
\left(\begin{array}{c} J_{ch} \\ J_{sp}\\J_Q \end{array}\right) =G_{ch}\left(\begin{array}{ccc}1&P&S_{ch}\\P&1&P'S_{ch}\\S_{ch}T&P'S_{ch}&\mathcal{K}/G_{ch}\end{array}\right)\left(\begin{array}{c}  -\mathcal{E}\\-\mathcal{E}_{sp} \\-\Delta T \end{array}\right)
\end{equation}
In the above Eq.(\ref{aa}), $\mathcal{E}$ is the applied electric field while the spin voltage applied $\mathcal{E}_{sp}=0$ in our system. Here we have summed the contribution of two valleys such that the total electric charge conductance $G_{ch}=G_\uparrow+G_\downarrow$ and spin conductance $G_{sp}=|G_{ch}P|$, with $G_s=\sum_vG_s^v$ and $S_s=\frac{1}{2}\sum_vS_s^v$ with $s=\uparrow,\downarrow$, $S_{ch}$ and $S_{sp}$ are the charge and spin Seebeck co-efficients respectively, $P$ is the polarization of spin conductance while $P'$ is the polarization of the product of Seebeck coefficient and conductance \cite{Bauer,bauer-rev} which are defined as follows:
\begin{eqnarray}\label{p'}
S_{ch}&=&\frac{G_\uparrow S_\uparrow+G_\downarrow S_\downarrow}{G_\uparrow +G_\downarrow},\quad
P=\frac{G_\uparrow-G_\downarrow}{G_\uparrow+G_\downarrow},\quad
P'=\frac{G_\uparrow S_\uparrow-G_\downarrow S_\downarrow}{G_\uparrow S_\uparrow+G_\downarrow S_\downarrow}.\nonumber\\
\end{eqnarray}
Similarly $S_{sp}=S_{ch}P'$. The spin polarization also affects the thermal conductance which is defined as-
\begin{eqnarray}\label{}
\mathcal{K}&=&\kappa+\frac{1+P^{'2}-2PP^{'}}{(1-P^2)}G_{ch}S_{ch}^2T,\\
\mbox{and on replacing $P,P'$ gives }\mathcal{K}&=&\kappa+G_\uparrow S_\uparrow^2T+G_\downarrow S_{\downarrow}^2T.
\end{eqnarray}
with $\kappa$ being the thermal conductivity in absence of any electrical charge or spin conductivity\cite{Bauer}, defined as-
\begin{eqnarray}\label{kappa} 
\kappa&=&\kappa_{\uparrow}+\kappa_{\downarrow},\qquad
\kappa_s=\frac{L_s^{11}L_s^{22}-L_s^{12}L_s^{21}}{L_s^{11}}, \mbox{ with $L_s^{ij}=\sum_vL_s^{ij, v}$ as in Eq.~(\ref{current})}.
\end{eqnarray}

\subsection{Efficiency and power of quantum spin heat engine}
 The charge(spin) power\cite{benetti} defined as usual as the product of electric current and voltage applied then can be written as-
\begin{eqnarray}\label{power}
\mathcal{P}_{ch(sp)}=J_{ch(sp)}\mathcal{E}=(G_{ch(sp)}\mathcal{E}+G_{ch}S_{ch(sp)}\Delta T)\mathcal{E}\nonumber\\
\end{eqnarray}
The above equation is maximized by $\frac{d\mathcal{P}_{ch(sp)}}{d\mathcal{E}}=0$, at $\mathcal{E}=-\frac{G_{ch}S_{ch(sp)}}{2G_{ch(sp)}}\Delta T$, which gives maximum power as-
\begin{eqnarray}\label{pmax}
\mathcal{P}_{ch}^{max}&=& \frac{1}{4}S_{ch}^2G_{ch}(\Delta T)^2, \mbox{   and     }
\mathcal{P}_{sp}^{max}=\frac{1}{4}S_{sp}^2\frac{G_{ch}^2}{G_{sp}}(\Delta T)^2
\end{eqnarray}
The efficiency at maximum power is defined as the ratio of maximum power to the heat current transported and can be derived as follows\cite{benetti}- 
\begin{eqnarray}\label{etapmax}
\eta(\mathcal{P}_{ch}^{max})&=& \frac{\mathcal{P}_{ch}^{max}}{J_Q}=\frac{\eta_c}{2}\frac{G_{ch}S_{ch}^2T/(\mathcal{K}-G_{ch}S_{ch}^2T)}{2+G_{ch}S_{ch}^2T/(\mathcal{K}-G_{ch}S_{ch}^2T)}\nonumber\\
&=&\frac{\eta_c}{2}\frac{ZT|_{ch}}{2+ZT|_{ch}}
\end{eqnarray}
\begin{eqnarray}\label{etapmaxsp}
\eta(\mathcal{P}_{sp}^{max})&=&\frac{\mathcal{P}_{sp}^{max}}{J_Q}=\frac{\eta_c}{2}\frac{G_{ch}^2S_{sp}^2T/(G_{sp}\mathcal{K}-G_{ch}^2S_{ch}S_{sp}T)}{2+G_{ch}^2S_{ch}S_{sp}T/(G_{sp}\mathcal{K}-G_{ch}^2S_{ch}S_{sp}T)}\nonumber\\
&=&\frac{\eta_c}{2}|P'|\frac{ZT|_{sp}}{2+ZT|_{sp}}
\end{eqnarray}
at $\mathcal{E}_{ch(sp)}=-\frac{G_{ch}S_{ch(sp)}}{2G_{ch(sp)}}\Delta T$, which is the condition for maximum power. Herein, $\eta_c$ is the Carnot efficiency defined by $\frac{\Delta T}{T}$ and $ZT|_{ch/sp}$ is the figure of merit, a dimensionless quantity, defined as-
\begin{eqnarray}\label{ZT}
ZT|_{ch}&=&\frac{G_{ch} S_{ch}^2 T}{\mathcal{K}-G_{ch}S_{ch}^2T},\\
ZT|_{sp}&=&\frac{P'G_{ch} S_{ch}^2 T}{P\mathcal{K}-P'G_{ch}S_{ch}^2T}.
\end{eqnarray}
Similarly, efficiency $\eta$ can be written as\cite{benetti}-
 \begin{eqnarray}\label{eta}
\eta_{ch}=\frac{\mathcal{P}_{ch}}{J_Q}=\frac{(G_{ch}\mathcal{E}+G_{ch}S_{ch}\Delta T)\mathcal{E}}{(G_{ch}S_{ch}T\mathcal{E}+\mathcal{K}\Delta T)},\\
\eta_{sp}=\frac{\mathcal{P}_{sp}}{J_Q}=\frac{(G_{sp}\mathcal{E}+G_{ch}S_{sp}\Delta T)\mathcal{E}}{(G_{ch}S_{ch}T\mathcal{E}+\mathcal{K}\Delta T)}.
\end{eqnarray}
To calculate maximal efficiency for the charge transported we need to calculate $\frac{d\eta_{ch}}{d\mathcal{E}}=0$ in Eq.~(\ref{eta}), this with the condition $J_Q>0$, gives-
\begin{eqnarray}
\mathcal{E}_{}&=&\frac{\mathcal{K}}{G_{ch}S_{ch}T}(-1+\sqrt{1-\frac{G_{ch}S_{ch}^2T}{\mathcal{K}}})\Delta T,\nonumber\\
\text{Thus,}\quad\eta_{ch}^{max}&=&\eta_c\frac{\sqrt{ZT|_{ch}+1}-1}{\sqrt{ZT|_{ch}+1}+1}.
\end{eqnarray}
Similarly, to calculate the maximal efficiency for spin transport we need to calculate $\frac{d\eta_{sp}}{d\mathcal{E}}=0$ in Eq.~(21), this again with the condition $J_Q>0$, gives-
\begin{eqnarray}
\mathcal{E}_{sp}&=&\frac{\mathcal{K}}{G_{ch}S_{ch}T}(-1+\sqrt{1-\frac{P'}{P}\frac{G_{ch}S_{ch}^2T}{\mathcal{K}}})\Delta T,\nonumber\\
\text{Thus,}\quad\eta_{sp}^{max}&=&\eta_c|P'|\frac{\sqrt{ZT|_{sp}+1}-1}{\sqrt{ZT|_{sp}+1}+1}.
\end{eqnarray}

After determining the expressions for the quantities (both charge as well as spin) like Seebeck coefficient, Thermoelectric figure of merit, maximum power output and efficiency of respective heat engines at maximum power, we plot them in section IV. We also discuss and analyze the aforesaid plots in the same section.
\section{Model}

\begin{figure}
\includegraphics[width=0.95\textwidth]{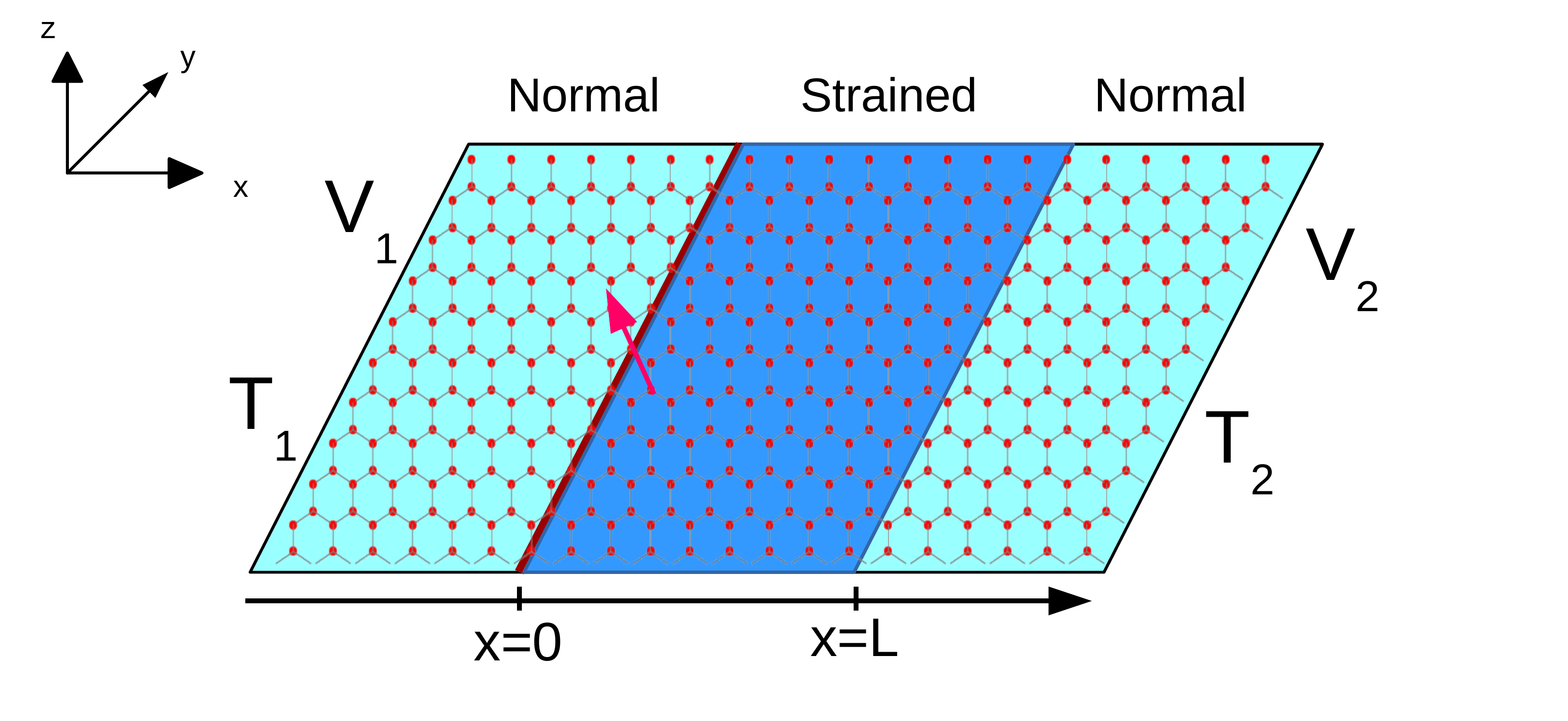}
\caption{ Monolayer graphene with a magnetic impurity at $x=0$ denoted by thick maroon line. The middle portion is strained region while the two side portions are normal graphene regions. Voltages $V_1$ and $V_2$ are applied to the two sides which are at temperatures $T_1$ and $T_2$ respectively.}
\end{figure}

\subsection{Hamiltonian}
\begin{figure}
      \centering \subfigure{\includegraphics[width=0.9\textwidth]{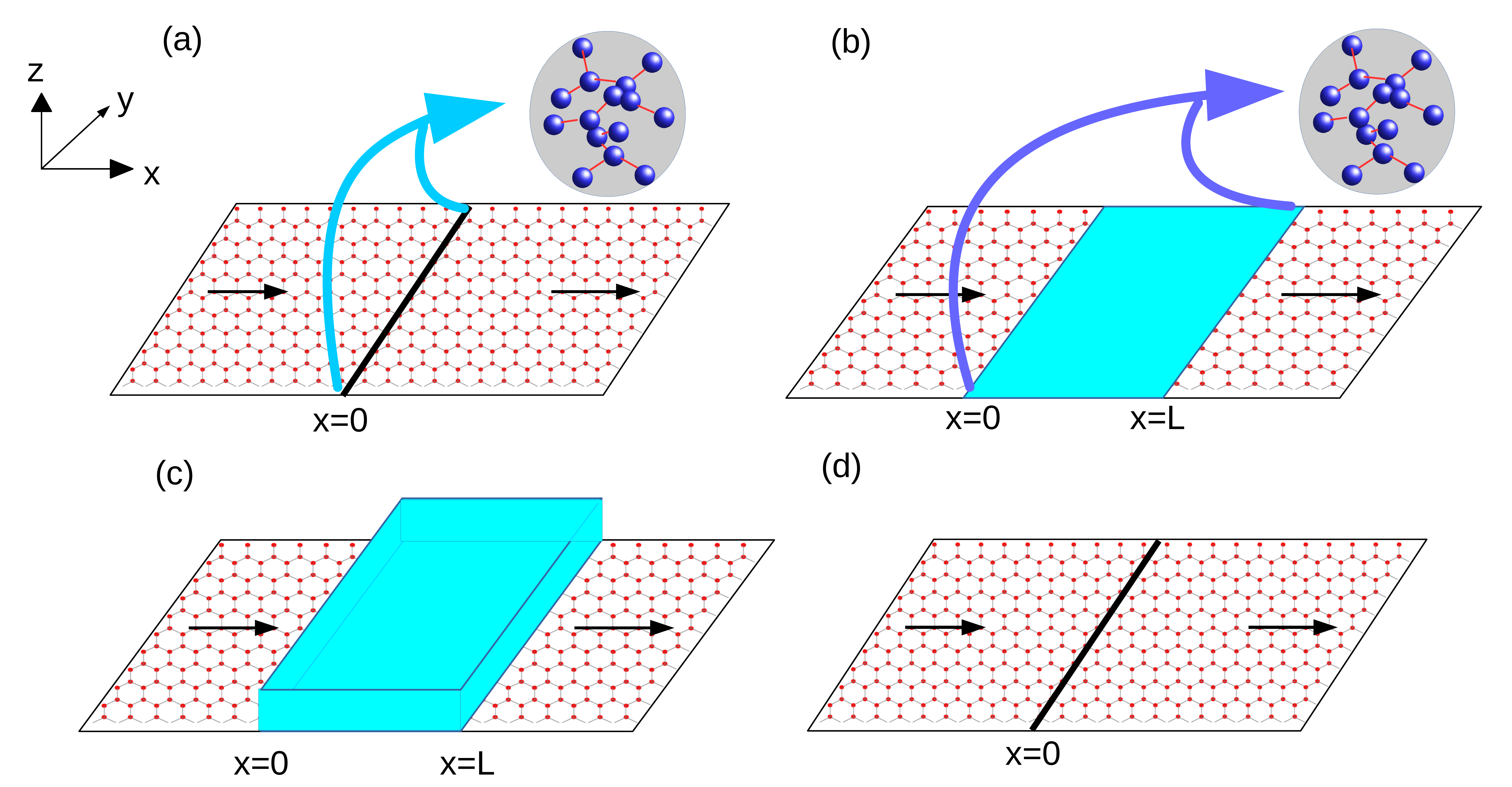}}
       \caption{2D graphene monolayer with (a) a delta potential magnetic impurity, (b)  rectangular barrier magnetic impurity, (c) a rectangular potential barrier and (d) a delta potential barrier. A rectangular barrier magnetic impurity(b) models a magnetic quantum dot (see Ref.\cite{maruri}) the transmission through which approximates that of a delta potential magnetic impurity(a) to a great extent. Similarly, a rectangular potential barrier(c) approximates a delta potential(d) in modeling the Klein paradox (see Refs.\cite{kat,bar}). }
       \label{exp}
\end{figure}

A graphene sheet is lying in the x-y plane, a strain is applied to the region $0<x<L$, see Fig.~1, with a magnetic impurity at $x=0$. The in-plane uniaxial strain impacts the hopping between nearest neighbors and is generally delineated via a gauge vector which takes opposing signs in the two valleys ($K$ and $K'$) of graphene \cite{castro}. In the Landau gauge, the vector potential
corresponding to the strain is $\vec A=(0, A_y)$. The system is then defined by the Hamiltonian-
\begin{equation}\label{36}
\mathcal{H_{K/K'}}=H_{K/K'}+J{\mathbf s.\mathbf S}\delta(x)
\end{equation}
with ${H}_K=\hbar v_f \sigma.(k-t')$ and $H_{K'}=\hbar v_f\sigma^*.(k+t')$, . Strain is denoted as $t=\hbar{v_f}t'= {A_y}[\Theta(x)-\Theta(x-L)]$ with $\Theta$ the Heaviside step function and $v_F$ the Fermi velocity. The first term in Eq.~(\ref{36}) represents the kinetic energy in graphene
with $\sigma=(\sigma_x, \sigma_y)$ - the Pauli matrices that operate on the sub-lattices
A or B and ${\bf k}=(k_x , k_y )$ the 2D wave vector. The second term in Eq.~(\ref{36}) denotes the exchange interaction between Dirac electron and magnetic impurity with $J$ representing the strength of the exchange interaction. The spin of Dirac electron is denoted by ${\bf s}$, while ${\bf S}$ represents spin of the magnetic impurity and $m$ its magnetic moment, while magnetic moment of Dirac electrons is $1/2$ (spin up) or $-1/2$ (spin down).
{For better understanding of our model we have compared our delta potential magnetic impurity with a rectangular  potential barrier magnetic impurity in Fig.~\ref{exp}. There is  a single magnetic impurity located along the line $x=0$. A solid black color line is shown at $x=0$ in Fig.~\ref{exp}(a). The magnetic impurity is lying along this line. The magnetic impurity is modeled as a delta potential in $x-$, but is uniform in the $y-$direction. A magnetic quantum dot doped with few $Mn^+$ ions can be thought of as a magnetic impurity, see Refs.~\cite{maruri, abolfath}. We assume it to have a finite width with a translational invariance in the $y-$direction. This can be understood with an analogy to a rectangular potential barrier in graphene. Klein tunneling in graphene is a 2D scattering problem, see Ref.~\cite{kat}. The Klein setup has a rectangular potential barrier between $x=0$ and $x=L$ with translational invariance in the $y-$direction, as shown in Fig.~\ref{exp}(c). The potential barrier affects the transmission of incident particles in the $x-$direction but doesn't affect the transmission in $y-$direction because the transmitting particle cannot feel the potential change in the $y-$direction. As one reduces the length $L$ of the potential barrier, it becomes similar to a delta potential located at $x=0$, see Ref.~\cite{bar}. Similarly, a magnetic impurity can have a finite width  between $x=0$ and $x=L$ with a translational invariance in the $y -$direction, as shown in Fig.~\ref{exp}(b), see Ref.~\cite{maruri}. If one decreases the width $L$ of the impurity, it reduces to a delta function like profile affecting the transmission in the $x-$direction but not in the $y-$direction, see Fig.~\ref{exp}(a). All of the electrons passing through the system interact with the impurity. Refs.~\cite{maruri, abolfath} have a magnetic impurity embedded into a graphene monolayer  very similar to us. The analysis as done in Refs.~\cite{firoz,firoz2,maruri}, is used in this work also. In Ref.~\cite{maruri}, a delta potential approximation of a rectangular barrier magnetic impurity in a graphene monolayer shows that for a range of incident angles from $-\pi/6$ to $\pi/6$ the difference between the transmissions through delta potential magnetic impurity and that of the rectangular barrier magnetic impurity is quite small. In Ref.~\cite{maruri} too, the delta potential magnetic impurity is an approximation for a magnetic quantum dot with spin.}

{ We consider a magnetic impurity as the prototype of a magnetic quantum dot doped with few $Mn^+$ ions, oriented by an external magnetic field and put in a specific state with spin $S$ and spin magnetic moment in z-direction $m$, see Refs.~\cite{maruri, abolfath}. It can be oriented such that only a particular state-defined by $S, m$ is occupied. Two types of scattering can happen: 1. with spin-flip (same $S$ but $m\rightarrow m\pm1$) or 2. without spin-flip (same $S$ as well as $m$) of magnetic impurity. The rest of the states would have zero occupation probability as shown in the analysis of the scattering of electrons due to the magnetic impurity in the next subsection, see also Refs.\cite{maruri,firoz,firoz2}.}

\subsection{Wave functions and boundary conditions}

To calculate the transmission probability and from it the Onsager coefficients and the thermoelectric factors we consider a spin-up electron with energy $E$ incident at the strained graphene interface at $x=0$ at an incident angle $\phi$ . At the interface itself we also have a magnetic impurity. The incident electron thus can be scattered due to the strained region. Further, its spin can also be affected because of the magnetic impurity. The incident electron thus can be scattered as a spin up or down electron depending on the spin and magnetic moment of the magnetic impurity.

The wave function for A-sub-lattice in each region (normal and strained) for K-
valley can be written as:\\
For $x<0$-
\begin{eqnarray}\label{a}
\psi_A^1(x,y)&=&\left[\begin{array}{c} (e^{ik_xx}+r_{\uparrow} e^{-ik_xx}) \chi_m \\r_{\downarrow} e^{-ik_xx} \chi_{m+1}\end{array}\right]\\
\psi_B^1(x,y)&=&\left[\begin{array}{c} (e^{ik_xx+i\phi}-r_{\uparrow} e^{-ik_xx-i\phi}) \chi_m \\-r_{\downarrow} e^{-ik_xx-i\phi} \chi_{m+1}\end{array}\right]
\end{eqnarray}
\\in region $0<x<L$-
\begin{eqnarray}\label{b}
\psi_A^2(x,y)&=&\left[\begin{array}{c} (a_{\uparrow} e^{iq_xx}+b_{\uparrow} e^{-iq_xx}) \chi_m\\ (a_{\downarrow} e^{iq_xx}+b_{\downarrow} e^{-iq_xx}) \chi_{m+1}\end{array}\right]\\
\psi_B^2(x,y)&=&\left[\begin{array}{c} (a_{\uparrow} e^{iq_xx+i\theta}-b_{\uparrow} e^{-iq_xx-i\theta}) \chi_m\\ (a_{\downarrow} e^{iq_xx+i\theta}-b_{\downarrow} e^{-iq_xx-i\theta})\chi_{m+1} \end{array}\right]
\end{eqnarray}
and for $x>L$-
\begin{eqnarray}\label{c}
\psi_A^3(x,y)&=&\left[\begin{array}{c} t_{\uparrow} e^{ik_xx} \chi_m \\t_{\downarrow} e^{ik_xx} \chi_{m+1}\end{array}\right]\\
\psi_B^3(x,y)&=&\left[\begin{array}{c} t_{\uparrow} e^{ik_xx+i\phi} \chi_m\\t_{\downarrow} e^{ik_xx+i\phi} \chi_{m+1}\end{array}\right]
\end{eqnarray}
\\
 The $x$ component of the wave-vector in strained region is $q_x=\sqrt{(E/\hbar v_F)^2-(k_y-t)^2}$, whereas in the normal region $q_x$ is substituted with $k_x$ , wherein $k_x=E \cos\phi/\hbar v_F$, and the phase factor in strained region is given by $\tan\theta=(k_y - t)/q_x$ . $\chi_m$ is the eigen state of $z-$component of spin operator of magnetic impurity $S_z$ with $S_z  \chi_m = m \chi_m$, $m$ being the corresponding eigen-value. The spin flipping mechanism is considered elastic and the sum of the $z-$components of the  spin magnetic moment of impurity($m$) and of electron($m'=\pm 1/2$), i.e., $M=m+m'$ remains conserved before and after spin-flip scattering. Following Ref.~\cite{helman}, one obtains the boundary conditions at $x =0$:
\begin{eqnarray}
i\hbar v_F[\psi_B^2(x=0)-\psi_B^1(x=0)]=\frac{J}{2}s.S[\psi_A^1(x=0)+\psi_A^2(x=0)]\nonumber\\
\end{eqnarray}
and
\begin{eqnarray}
i\hbar v_F[\psi_A^2(x=0)-\psi_A^1(x=0)]=\frac{J}{2}s.S[\psi_B^1(x=0)+\psi_B^2(x=0)]\nonumber\\
\end{eqnarray}
and at $x=L$ as-
\begin{eqnarray}
\psi_A^2(x=L)=\psi_A^3(x=L)
\end{eqnarray}
and
\begin{eqnarray}
\psi_B^2(x=L)=\psi_B^3(x=L)
\end{eqnarray}

The spin flip process is attributed to the interaction between the spin of electron $({\bf s})$ and the spin of magnetic impurity $(\bf S)$, with
${\bf s.S}=s_zS_z+\frac{1}{2}(s^-S^++s^+S^-)$, where $s^-S^+\left[\begin{array}{c} 1 \\0\end{array}\right]\chi_m=F\left[\begin{array}{c} 0 \\1\end{array}\right]\chi_{m+1}$ and $s^+S^-\left[\begin{array}{c} 0 \\1\end{array}\right]\chi_m=F'\left[\begin{array}{c} 1 \\0\end{array}\right]\chi_{m-1}$ with $F=\sqrt{(S-m)(S+m+1)}$ and $F'=\sqrt{(S+m)(S-m+1)}$. Here, $s_z$ with $s_z\left[\begin{array}{c}1\\0\end{array}\right]=\frac{1}{2}\left[\begin{array}{c}1\\0\end{array}\right]$ and $S_z$ are the z-components of
the spin operator of electron and magnetic impurity, respectively. $S_{\pm} = S_x \pm iS_y$, where $S_+$ and $S_-$ are the spin raising and spin lowering operators for magnetic impurity, and $s_{\pm} =s_x \pm is_y$ are the same for electrons.

After substituting the wave functions (30-35) in Eqs.~(36-39), at $x=0$ we get- 
\begin{eqnarray}\label{40}
&a_{\uparrow}&(e^{i\theta}+i\alpha m)-b_{\uparrow}(e^{-i\theta}-i\alpha m)-(e^{i\phi}-i\alpha m)\nonumber\\&+&r_{\uparrow}(e^{-i\phi}+i\alpha m)+i\alpha F(a_{\downarrow}+b_{\downarrow}+r_{\downarrow})=0,\\
&a_{\downarrow}&(e^{i\theta}-i\alpha (m+1))-b_{\downarrow}(e^{-i\theta}+i\alpha(m+1))\nonumber\\&+&r_{\downarrow}(e^{-i\phi}-i\alpha(m+1))+i\alpha F(a_{\uparrow}+b_{\uparrow}+r_{\uparrow}+1)=0,\\
&a_{\uparrow}&(1+i\alpha me^{i\theta})+b_{\uparrow}(1-i\alpha me^{-i\theta})-(1-i\alpha me^{i\phi})\nonumber\\&-&r_{\uparrow}(1+i\alpha me^{-i\phi})+i\alpha F(a_{\downarrow}e^{i\theta}-b_{\downarrow}e^{-i\theta}-r_{\downarrow}e^{-i\phi})=0,\\
&a_{\downarrow}&(1-i\alpha (m+1)e^{i\theta})+b_{\downarrow}(1+i\alpha(m+1)e^(-i\theta))-r_{\downarrow}(1\nonumber\\&-&i\alpha(m+1)e^{-i\phi})+i\alpha F(a_{\uparrow}e^{i\theta}-b_{\uparrow}e^{-i\theta}+e^{i\phi}-r_{\uparrow}e^{-i\phi})=0,\nonumber\\
\end{eqnarray}
and at $x=L$ we get-
\begin{eqnarray}
t_{\uparrow}e^{ikL}&=&a_{\uparrow}e^{iqL}+b_{\uparrow}e^{-iqL}\\
t_{\downarrow}e^{ikL}&=&a_{\downarrow}e^{iqL}+b_{\downarrow}e^{-iqL}\\
t_{\uparrow}e^{ikL+i\phi}&=&a_{\uparrow}e^{iqL+i\theta}-b_{\uparrow}e^{-iqL-i\theta}\\
t_{\downarrow}e^{ikL+i\phi}&=&a_{\downarrow}e^{iqL+i\theta}-b_{\downarrow}e^{-iqL-i\theta}
\end{eqnarray}
In the above equations $\alpha=J/(4 \hbar v_f)$. Eqs.~(40)-(47) consist of 8 unknowns which satisfy the probability conservation- $|r_{\uparrow}|^2+|t_{\uparrow}|^2+|r_{\downarrow}|^2+|t_{\downarrow}|^2=1$. Similarly, for spin down incident electron from the left side we can derive the scattering amplitudes. Further, for $K'$ valley too solving the Hamiltonian one can get the transmission amplitude $(t_s)$ and reflection amplitude $(r_s)$ with $s={\uparrow, \downarrow}$ again in a nod to probability conservation satisfying $|r_{\uparrow}|^2+|t_{\uparrow}|^2+|r_{\downarrow}|^2+|t_{\downarrow}|^2=1$ for $K'$ valley also. Since there is no inter valley scattering, our results  remain identical for $K'$ valley after integrating over both energy and the incident angle. So we focus on the transmission probability $\mathcal{T}_{s} = |t_s|^2 $ in one valley $v=K$ only (see Eq.~8), in the results and discussion section IV.

\begin{figure}
  \centering{\includegraphics[width=0.95\textwidth]{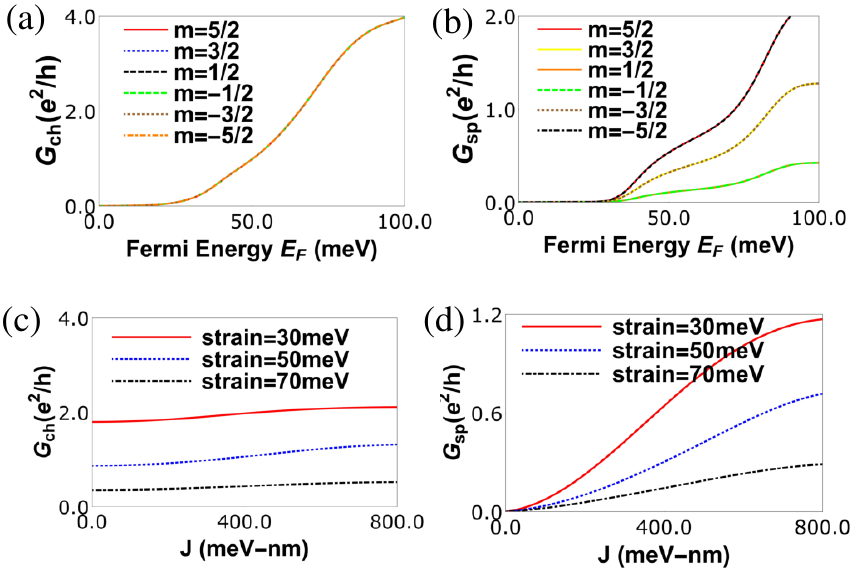}}  
       \caption{(a) Charge Conductance ($G_{ch}$) vs. $E_F$ (Fermi energy) for various values of magnetic moment $m$, length of strained graphene layer $L=40$ nm and width  $W=20$ nm, strain $t=50$ meV, temperature $T=30$ K with spin of magnetic impurity $S=5/2$ and $J=-600$ meV-nm, (b) Spin Conductance ($G_{sp}$) vs. $E_F$ (Fermi energy) for various values of magnetic moment $m$, length of strained graphene layer $L=40$ nm, strain $t=50$ meV, temperature $T=30$ K with spin of magnetic impurity $S=5/2$ and $J=-600$ meV-nm, (c) Charge conductance ($G_{ch}$) vs $J$ (impurity coupling strength) for various strains at Fermi energy $E_F=50$ meV, length of strained graphene layer $L=60$ nm, temperature $T=30$ K with spin of magnetic impurity $S=5/2$ and spin magnetic moment $m=-5/2$. (d)Spin conductance ($G_{sp}$) vs. $J$ (impurity coupling strength) for various strains at Fermi energy $E_F=50$ meV, length of strained graphene layer $L=60$ nm, temperature $T=30$ K with spin of magnetic impurity $S=5/2$ and spin magnetic moment $m=-5/2$. }
       \label{conductance}
\end{figure}
 \begin{figure}
      \centering{\includegraphics[width=0.97\textwidth]{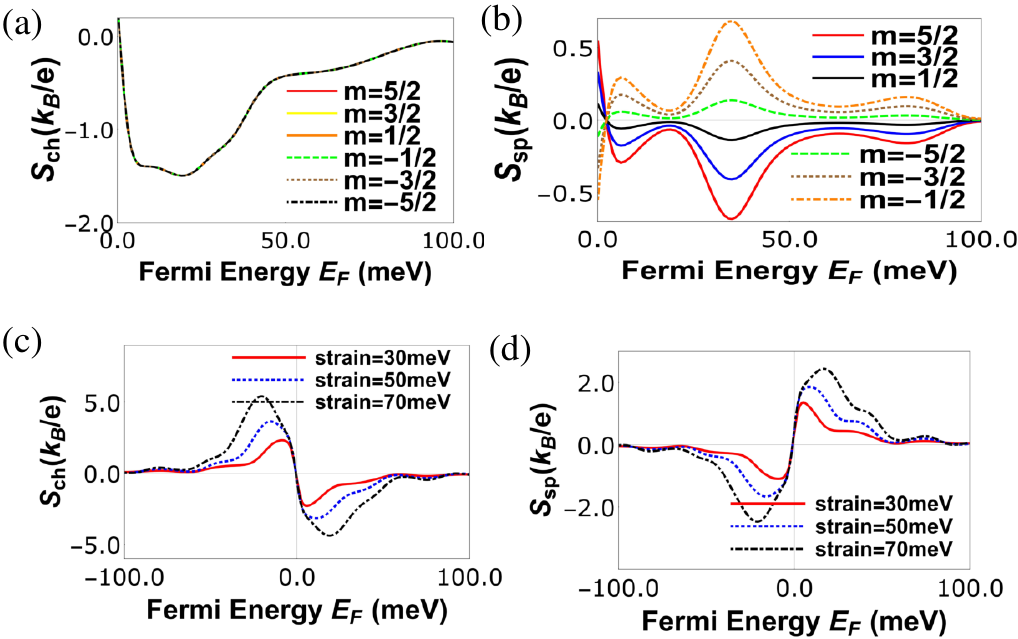}}
       \caption{(a) Charge Seebeck coefficient ($S_{ch}$) vs. Fermi energy for various $m$ of magnetic impurity at $T=30$ K, $J=-600$ meV-nm, strain $(t)= 50$ meV and spin $S=5/2$ and length of strained graphene region $L=40$ nm and width $w=20$ nm and (b) Spin Seebeck coefficient $S_{sp}$ vs. Fermi energy $E_F$ in meV for various $m$ of magnetic impurity, length of strained graphene layer $L=40$ nm, strain $=50$ meV, temperature $T=30$ K with spin of magnetic impurity $S=5/2$ and $J=-600$ meV-nm. (c) Charge Seebeck coefficient ($S_{ch}$) vs Fermi energy for various strains at $J=600$ meV-nm, length of strained graphene layer $L=60$ nm, temperature $T=30$ K with spin of magnetic impurity $S=5/2$ and spin magnetic moment $m=-5/2$, (d) Spin Seebeck coefficient ($S_{sp}$) vs Fermi energy for various strains with parameters same as (c).}
       \label{Seebeck}
\end{figure}
  \begin{figure}
      \centering \subfigure{\includegraphics[width=0.95\textwidth]{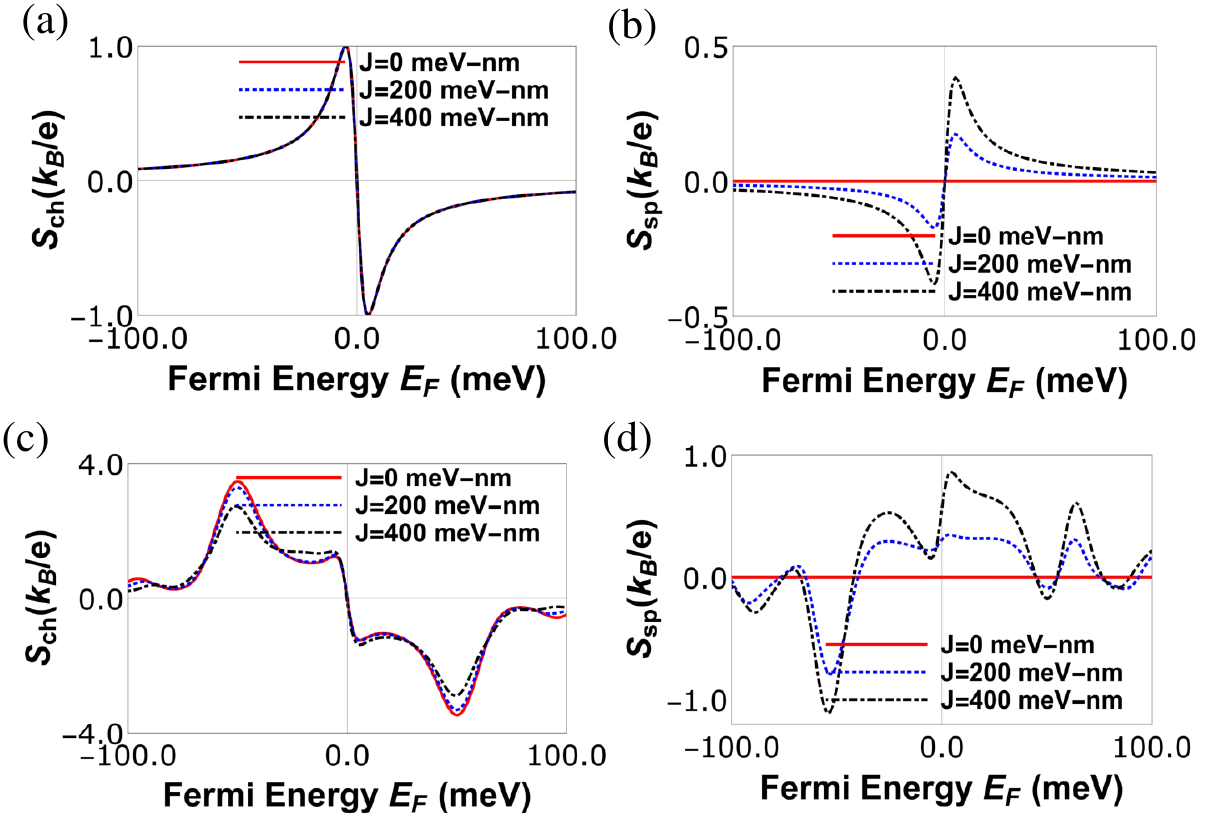}}
       \caption{(a) Charge Seebeck coefficient $S_{ch}$ vs. Fermi energy ($E_F$) for various exchange coupling strength $J$ with parameters at $L=40$ nm, strain $t=0$ meV, $T=30$ K, $S=5/2$, $m=-5/2$, (b) spin Seebeck coefficient $S_{sp}$ vs. Fermi energy ($E_F$) for various exchange coupling strength $J$ with parameters at $L=40$ nm, strain $t=0$ meV, $T=30$ K, $S=5/2$, $m=-5/2$, (c) charge Seebeck coefficient $S_{ch}$ vs. Fermi energy ($E_F$) for various exchange coupling strength $J$ with parameters at $L=40$ nm, strain $t=100$ meV, $T=30$ K, $S=5/2$, $m=-5/2$, (d) spin Seebeck coefficient $S_{sp}$ vs. Fermi energy ($E_F$) for various exchange coupling strength $J$ with parameters at $L=40$ nm, strain $t=100$ meV, $T=30$ K, $S=5/2$, $m=-5/2$.}
       \label{Seebeckvsj}
\end{figure}
\section{Results and Discussion}
 \subsection{Charge/spin conductance and charge/spin Seebeck coefficient} 
 In Figs.~\ref{conductance} (a) and (b) we plot the charge and spin conductance for various $m$ values (spin magnetic moment in z-direction) of magnetic impurity. We see that though different magnetic orientations have no effect on the charge conductance, the spin conductance increases as the magnitude of $m$ increases, but it is unaffected by the direction of $m$. 
 {{ Decreasing the spin magnetic moment $m$ of the magnetic impurity reduces the transmission probability of spin up electrons, but increases the transmission probability for spin down electrons by the same amount. Thus the sum of transmission probabilities of spin-up and spin-down electrons remains unchanged with decreasing $m$.} The total charge conductance remains unaffected by the changing $m$, but the spin conductance increases as the difference between spin up current and spin down current increases. {It can also be noted that both charge as well as spin conduction gaps shown in Figs.~\ref{conductance}(a, b) is not due to the band gap in the band structure but due to the shift of Dirac cones in the Brillouin zone in strained region of monolayer graphene. In presence of strain, Dirac cones are formed/shifted in strained region of graphene at different positions in the $k_x$ axis while in the unstrained region of graphene there is no shift. Thus, there is a energy gap created in the strained region of graphene. The conduction band is related to the energy gap in $k_x$ axis. Thus, there is always a finite transmission gap obtained even for small strain value in our device and shown in Figs.~\ref{conductance}(a, b). This conduction gap can be of two types- charge and spin conduction gap for our device. The charge and spin conduction gaps are always proportional to the strength of strain. This result is one of the reasons to study the superior thermoelectric effects in graphene nano structure in presence of strain. It is to be noted here that the presence of exchange interaction does not affect the charge conduction gap while it reduces the spin conduction gap.} Similar effects on the charge/spin conductances are observed when  the exchange interaction $J$ is altered.} { The effect of the exchange interaction $J$ on transmission probability of incident spin up and spin down electrons is same as the effect of magnetic moment $m$ of the magnetic impurity on it.} In Fig.~\ref{conductance} (c) we see that the charge conductance is almost constant as function of the exchange interaction ($J$), however the spin conductance increases as shown in Fig.~\ref{conductance} (d). { This can be understood in this way- Increasing exchange coupling $J$ of the magnetic impurity increases the transmission probability of either spin-up (or, spin-down) electron depending on the magnetic moment $m$ of the magnetic impurity and reduces it for the spin-down (or, spin-up) electron. In this way the sum of the transmission probabilities of spin-up and spin-down electrons remains unchanged and so does the charge conductance. However, the difference between the transmission probabilities of spin-up and spin-down electrons increases showing a increase in spin conductance with exchange coupling $J$.} { If an electron is incident at the interface of strained and unstrained region, it is refracted to the strained region with a refraction angle $\theta=tan^{-1}{(k_y-t)/q_x}$ in $K$ valley. So, if one increases the strain $t$, electrons with incident angle $0$ to $\pi/2$ will refract close to the normal to the interface between the two regions and thus their transmission probability increases, but electrons with incident angle $0$ to $(-\pi/2)$ will refract away from the normal to interface reducing the transmission probability more and thus reducing the overall transmission (after integrating over incident angle $\phi$) in the the $K$ valley. In the $K'$ valley, the electrons refract in the opposite direction to that of the $K$ valley with a refraction angle $\theta=tan^{-1}{(k_y+t)/q_x}$, but overall transmission probability (after integrating over incident angle $\phi$) reduces with strain and is always equal to the $K$ valley unless a magnetic field is applied at the interface to create a valley polarization, see Ref. \cite{can}}. Increasing strain decreases both the charge as well as spin conductances. Similar to Fig.~\ref{conductance}, in Fig.~\ref{Seebeck} we see the effect of  the orientation of the magnetic impurity in $z-$ direction ($m$) and 
strain on the charge and spin Seebeck coefficients. In Figs.~\ref{Seebeck} (a) and (b) we see that impurity orientation $m$ has no effect on the charge Seebeck coefficient, but it has a huge impact on the spin Seebeck coefficient. { This can be understood as follows. With changing sign of the spin magnetic moment $m$ of the magnetic impurity from positive to negative, the transmission probability of spin-up electrons decreases while that for spin-down electrons increases by the same amount. Thus the total transmission probability of spin-up and spin-down electrons remains same with decreasing $m$. So, the charge Seebeck coefficient remains unaffected with changing $m$ but spin Seebeck coefficient decreases and changes its sign from positive to negative as seen in Figs.~\ref{Seebeck}(a, b).} In Figs.~\ref{Seebeck} (c) and (d) we see that charge and spin Seebeck coefficients both increase with increasing strain, which is opposite to the effect on charge and spin conductances. { This can be understood as follows- A bandgap in a nanostructured material can increase the Seebeck coefficient significantly. In graphene, due to its gapless band-structure the Seebeck coefficient is very small, see Ref.~\cite{dolphas}. Applying strain in a graphene device can shift the Dirac points in opposite direction by opening a conduction gap without opening a bandgap. This conduction gap increases with increasing strain and so also the charge/spin Seebeck coefficients.}

From Fig.~\ref{Seebeck} (b), it's evident that spin Seebeck coefficient depends on the sign (orientation) of the magnetic impurity $m$, i.e., $S_{sp}|_{m}=-S_{sp}|_{-m}$, unlike the spin conductance which is independent, since $G_{sp}=|G_{\uparrow}-G_{\downarrow}|$.  In Fig.~\ref{Seebeckvsj}(a) we see that exchange interaction strength $J$ has no effect on charge Seebeck coefficient $S_{ch}$ at zero strain, while the spin Seebeck coefficient $S_{sp}$ increases with $J$, as shown in Fig.~\ref{Seebeckvsj}(b). In presence of strain, the effect of $J$ on $S_{ch}$ is negligible.  One thing to note in Figs.~\ref{Seebeck} and \ref{Seebeckvsj} is that both spin as well as charge Seebeck coefficients are anti-symmetric as function of Fermi energy $(E_F)$, i.e., $S_{ch/sp}(E_F)=-S_{ch/sp}(-E_F)$ at zero strain. In presence of finite strain while $S_{ch}(E_F)=-S_{ch}(-E_F)$, $S_{sp}$ has no symmetry with respect to sign reversal of Fermi energy, in effect change of charge carriers from electrons to holes. All this is in contrast to the spin and charge conductances which are symmetric, $G_{ch/sp}(E_F)=G_{ch/sp}(-E_F)$, to reversal of charge carriers.
 
The sign change seen in Fig.~\ref{Seebeckvsj}(a) for the charge Seebeck coefficient $S_{ch}$ near the charge neutrality point or Dirac point is because the charge carriers switch from electrons to holes. 
 The origin of second peak in Fig.~\ref{Seebeckvsj}(c) is solely strain. On the other hand the first peak seen in Fig.~\ref{Seebeckvsj}(c)  which appears close to the Dirac point is due to the asymmetric contribution to the Seebeck current from electrons and holes, which arises due to the unique energy dependent density of states of graphene. The first peak is always present in graphene even in absence of strain, see Figs.~\ref{Seebeckvsj}(a) and (b) and Ref.~\cite{zuev}. In presence of strain, in addition to this unique energy dependent density of states of  graphene, an asymmetry is created in the transmission probability as function of energy and that gives rise to the second peak in Fig.~\ref{Seebeckvsj}(c). See also Ref.~\cite{arjun}, where a similar peak is observed due to strain in graphene. It should be noted that the position of the first peak is always fixed, i.e., close to the Dirac point but changing the parameters like length of the strained region one can change the position of the second peak and thus these two peaks may merge to form a single large peak, see Figs.~\ref{Seebeck}(c) and (d) which in turn leads to large power and efficiency.  It is to be  noted from Figs.~\ref{Seebeckvsj} (c) and (d) that at the Dirac point the charge Seebeck coefficient is exactly zero, while the spin Seebeck coefficient is finite, leading to the generation of pure spin current within the system due to temperature difference only. {In our paper, strain and magnetic impurity are used to enhance the charge and spin thermoelectric
properties respectively. Increasing strain increases the scattering of the electron while reducing the electrical conductance and thermal conductance as well. However, this increase in scattering due to strain also increases the charge Seebeck coefficient, which is inversely
proportional to the electrical conductance. On the other hand the splitting of the spin bands due to the magnetic impurity helps in generating large spin Seebeck coefficient. Since charge and spin thermoelectric properties are proportional to the square of the charge and spin Seebeck coefficient, they increase with increasing Seebeck coefficient. In this way, strain and magnetic impurity help in enhancing the performance of our graphene spin heat engine.}

Finally, we have neglected the phonon contribution in our calculations since the phonon contribution to the thermal conductance of graphene is quite small (almost absent) at low temperatures $0-30 K$, see Figs.~2,3  on Ref.~\cite{mazamutto} and Fig.~5 of Ref.~\cite{yong}. Beyond $25-30 K$ range, the phonon contribution increases linearly with temperature, as shown in Refs.~\cite{mazamutto,yong}. Thus, the phonon contribution to the thermal conductance can be neglected at the temperature range $20-30 K$ discussed in our work. 

  \subsection{Thermoelectric figure of merit, power and efficiency of quantum charge/spin heat engine}
To get large efficiency for our charge and spin heat engines we need a large charge and spin thermoelectric figure of merit ($ZT|_{ch}$ and $ZT|_{sp}$). From Eq.~(\ref{ZT}) we see that charge thermoelectric figure of merit is proportional to the product of square of the charge Seebeck coefficient $S_{ch}$ and charge conductance $G_{ch}$, i.e., $S_{ch}^2G_{ch}$, while spin thermoelectric figure of merit ($ZT|_{sp}$) is proportional to the product of charge and spin Seebeck ($S_{ch}$ and $S_{sp}$) coefficients with charge conductance $G_{ch}$, i.e., $S_{ch}S_{sp}G_{ch}=P'G_{ch}S_{ch}^2$ as in Eq.~(19). {Thus spin thermoelectric figure of merit $ZT|_{sp}$ is proportional to the polarization of the product of conductance ($G$) and Seebeck coefficient ($S$), i.e., $P'$ and inversely proportional to the polarization of electrons, $P$, see Eq.~(19). To get a large charge thermoelectric figure of merit $ZT|_{ch}$ we need a large charge Seebeck coefficient ($S_{ch}$) and large electrical charge conductance ($G_{ch}$) of the electrons while for spin thermoelectric figure of merit, we need a large spin Seebeck coefficient ($S_{sp}$) and small spin conductances ($G_{sp}=PG_{ch}$) or small polarization $P$ of the electrons. }
 
\begin{figure}
      \centering{\includegraphics[width=0.95\textwidth]{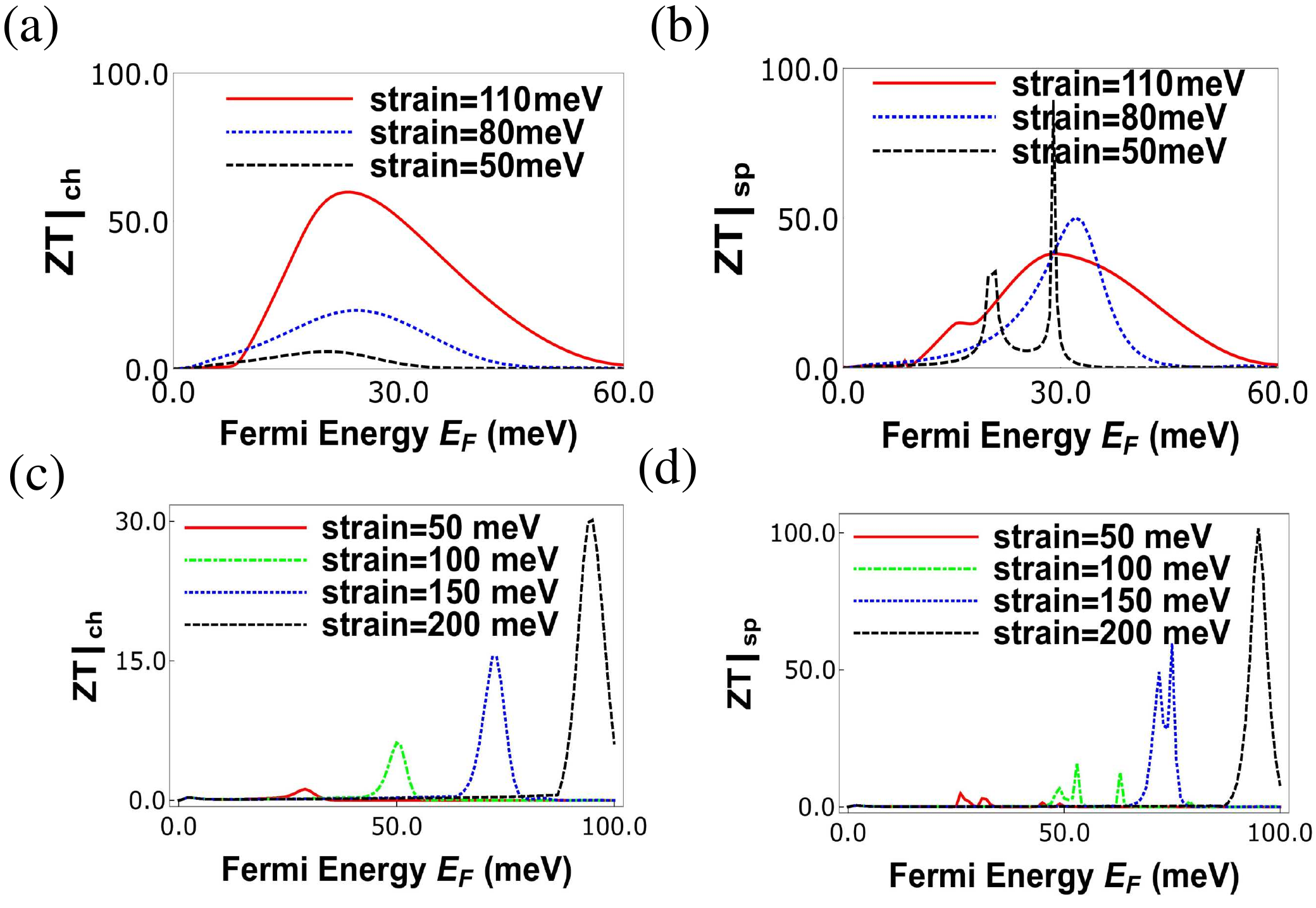}}
       \caption{(a) Charge thermoelectric figure of merit ($ZT|_{ch}$) vs. Fermi energy for various strains with parameters $T=30$ K, $J=232$ meV-nm and $L=80$ nm, $w=20$ nm and spin $S=5/2$, magnetic moment $m=-5/2$ of magnetic impurity, (b) Spin thermoelectric figure of merit $ZT|_{sp}$ vs. Fermi energy for various strains with parameters $T=30$ K, $J=232$ meV-nm and $L=80$ nm, $w=20$ nm and spin $S=5/2$, magnetic moment $m=-5/2$ of magnetic impurity. (c) $ZT|_{ch}$, (d) $ZT|_{sp}$ vs. Fermi energy at $T=10$ K, with all the other parameters same as in 6(a).}
     \label{zt}
\end{figure}

 In Fig.~\ref{zt}, charge and spin thermoelectric figure of merits are plotted as function of Fermi energy for various strains. In Fig.~\ref{zt} (a) we see that charge figure of merit $ZT|_{ch}$ increases with strain, while spin figure of merit $ZT_{sp}$ decreases as shown in Fig.~\ref{zt} (b). {Since increasing strain increases the charge Seebeck coefficient ($S_{ch}$) and decreases the electrical charge conductance ($G_{ch}$), so as a result $ZT|_{ch}$ increases with strain (as it is proportional to the square of the charge Seebeck coefficient). On the other hand for spin transport spin Seebeck coefficient increases with increasing strain while spin conductance decreases ($G_{sp}=PG_{ch}$) such that a large $ZT|_{sp}$ results. $ZT|_{sp}$ increases with strain in the overall picture however Fermi energy where maximum peak is observed is at low strain due to dependencies on other parameters.} $ZT|_{ch}$ takes values around $50$ which is quite large and similar to those obtained in Ref.~\cite{Bauer}. Further, $ZT|_{sp}$ approaches $100$ which is completely unheard of. These giant charge and spin thermoelectric factors are crucial for designing highly efficient quantum charge and spin heat engines and are one of the main novelties of this work. {This large charge figure of merit can be explained as follows. We have defined the figure of merit in our paper as $ZT|_{ch}=\frac{G_{ch}T}{\kappa}S_{ch}^2$. According to Wiedemann-Franz law, $\frac{G_{ch}T}{\kappa}=\frac{1}{L_0}$,where $L_0= $ is the Lorentz number which is a constant. So, $ZT|_{ch}$ is $S_{ch}^2/L_0$, and completely depends on the square of the charge Seebeck coefficient. We have shown in Fig.~(4) of our paper that both  charge as well as spin Seebeck coefficients increase monotonically with strain. If we increase strain in graphene, it creates a conduction gap by shifting the two Dirac cones in exactly opposite directions, and thus increases the Seebeck coefficient. Due to the lack of any band gap in pristine graphene, its Seebeck coefficient is very small and finite. So, if we keep on increasing the strain ($t=110 meV=4\% strain$), a large Seebeck coefficient, as large as $14(k_B/e)$(this value is two times the value of Seebeck coefficient observed in Fig.~2(b) of Ref.~16 of our paper with $4\%$ strain) can be generated. So, if we divide the square of Seebeck coefficient by the Lorentz number ($L_0=2.44*10^{-8}$ in SI unit) we will get $ZT|_{ch}\sim60$. Thus the large charge/spin Seebeck coefficients seen in presence of strain is the sole reason to achieve the huge $ZT|_{ch}$. Similarly, we can explain the huge $ZT|_{sp}$ as due to the large spin Seebeck coefficient. }
 
  According to Eq.~\ref{etapmax}, this large $ZT|_{ch}$ will give rise to a large efficiency at maximum power, $\eta(P_{ch}^{max})=0.48 \eta_C$, corresponding to this value of $ZT_{ch}$ the maximum charge power delivered by our QSHE is $0.02 (k_b\Delta T)^2/h$, which is quite small. This is the remarkable trade off between power and efficiency in a quantum heat engine, that when efficiency is maximum the corresponding power is minimum. That's why, we choose a set of parameters where power and efficiency both are moderately large to give the optimal performance. With a certain set of parameters, we obtain $ZT|_{ch}\sim 2$ for which we get the efficiency at maximum charge power $\eta(P_{ch}^{max}) = 0.166 \eta_C$ and maximum power delivered $\equiv 0.16 \frac{(k_B\Delta T)^2}{h}$, which is a large value compared with some other charge QHE's, see Table~1 (section VI). {In Fig.~6(c) and (d), we see that the charge and spin thermoelectric figure of merit is large ($\sim 30$) even at $T=10$ K, however, this large $ZT|_{ch/sp}$ value is reached only when the strain is large. Here, for calculating the charge/spin figure of merit we have neglected the phonon contribution to the thermal conductance. At temperature $t=30$ K, this phonon contribution is small but still can reduce the $ZT|_{ch/sp}$, if included in the calculation. However, at $T=30$ K, when the phonon contribution is very small and can be neglected, we see that $ZT|_{ch/sp}$ are still large ($\sim30$). Here, it is to be noted that at small temperatures ($T\sim10$ K) the figure of merit and the efficiency at maximum power remain huge at higher strain, although, the maximum power at this parameters reduces to very small values. That is why we have chosen $T=30$ K temperature as the optimum temperature of our model.}   

   \begin{figure}
      \centering {\includegraphics[width=0.99\textwidth]{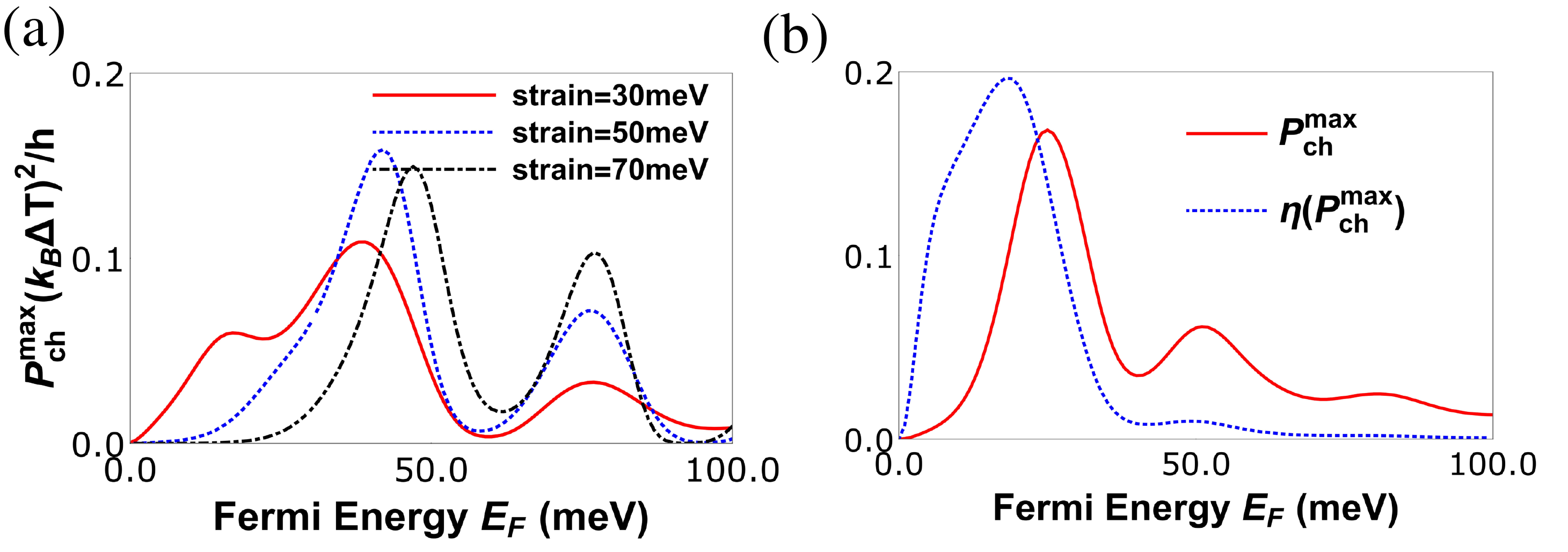}}
       \caption{(a) Maximum power for charge  current ($P_{ch}^{max}$) vs. Fermi energy ($E_F$) in meV for various strains at $J=600$ meV-nm, length of strained graphene layer $L=60$ nm, width $W=20$ nm, temperature $T=30$ K with spin of magnetic impurity $S=5/2$ and spin magnetic moment $m=-5/2$, (b) Maximum charge power ($P_{ch}^{max}$) and efficiency at maximum power ($\eta(P_{ch}^{max})$) vs. Fermi energy ($E_{F}$) in meV for  strain $t=30$ meV, $L=70$ nm, $W=20$ nm, $T=30$ K, $J=232$ meV-nm and spin $S=5/2$, magnetic moment $m=-5/2$ of magnetic impurity. }
       \label{powerch}
\end{figure}
  \begin{figure}
      \centering \subfigure{\includegraphics[width=0.95\textwidth]{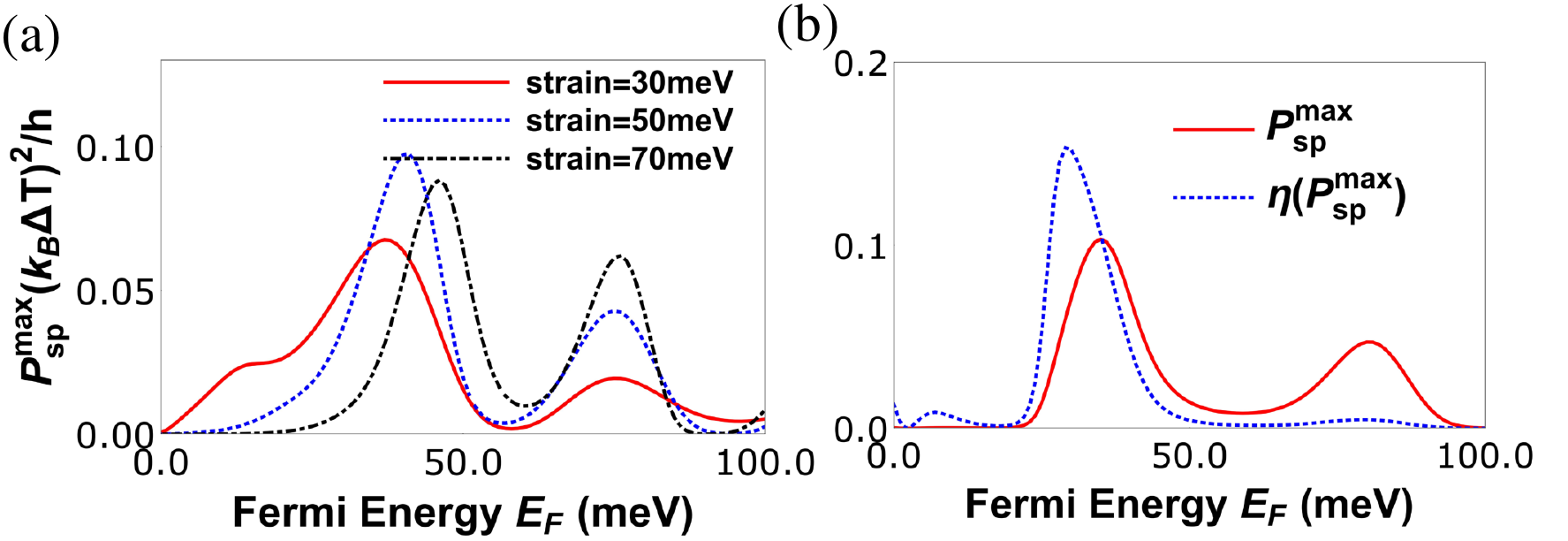}}
       \caption{(a) Maximum power for spin current ($P_{sp}^{max}$) vs. Fermi energy ($E_F$) for various strains with parameters $J=600$ meV-nm, $L=60$ nm, $W=20$ nm, $T=30$ K, $S=5/2$, $m=-5/2$, (b) Maximum spin power ($P_{sp}^{max}$) and efficiency at maximum power ($\eta(P_{sp}^{max})$) in units of $\eta_c$ vs. Fermi energy ($E_F$) at $L=40$ nm, $W=20$ nm, $T=30$ K, $J=-600$ meV-nm, strain $t=50$ meV, $S=5/2$, $m=-5/2$. }
       \label{powersp}
\end{figure}
In Fig.~\ref{powerch} we plot the maximum power for charge heat engine at various strains, we see that there are two peaks in $P_{ch}^{max}$. The first peak in $P_{ch}^{max}$ (which is proportional to $S_{ch}^2G_{ch}$) is observed when the charge conductance $G_{ch}$ dominates over the charge Seebeck coefficient $S_{ch}$, which can be seen at strain ($t=50$ meV). The second peak appears when the charge Seebeck coefficient $S_{ch}$ dominates over the charge conductance $G_{ch}$, this can be verified easily because the second peak increases with increasing strain. In Fig.~\ref{powerch}(b) we plot both maximum power($P_{ch}^{max}$) and the efficiency at maximum power ($\eta(P_{ch}^{max})$) as function of the Fermi energy ($E_F$). We see that $\eta(P_{ch}^{max})$ goes to almost $0.2\eta_c$, this is also a very large value as compared to other similar heat engines. The efficiency at maximum charge power as derived from Eq.~(\ref{etapmax}) depends only on $ZT|_{ch}$. Since in our case $ZT|_{ch}$ takes quite high values its not surprising that we have a highly efficient charge heat engine. Further, we see that the efficiency $\eta(P_{ch}^{max})$ is maximum ($0.2 \eta_c$) for $E_F=18 meV$ but at this Fermi energy the maximum power delivered is around $0.1 (k_B \Delta T)^2/h$. However, at Fermi energy close to $23.5 meV$ the efficiency although slightly lower at $0.16 \eta_c$ the maximum power output is $0.16  (k_B \Delta T)^2/h$. We not only need high efficiency but we need to deliver large output power too,  balancing these two needs implies operating the charge heat engine at $E_{F}=23.5 meV$ will satisfy both our needs. Similarly, in Fig.~\ref{powersp} we plot the maximum power for spin heat engine for various strains, we see that there are two peaks in $P_{sp}^{max}$ also. The first peak in $P_{sp}^{max}$ (which is proportional to $S_{sp}^2\frac{G_{ch}^2}{G_{sp}}$) is observed when the factor $\frac{G_{ch}^2}{G_{sp}}$ dominates over the spin Seebeck coefficient $S_{sp}$, which can be seen at strain $t=50 meV$. The second peak appears when the spin Seebeck coefficient $S_{sp}$ dominates over the factor $\frac{G_{ch}^2}{G_{sp}}$, this can be again verified as the second peak increases with increasing strain. In Fig.~\ref{powersp}(b) we plot both maximum power($P_{sp}^{max}$) and efficiency at maximum power ($\eta(P_{sp}^{max})$) as function of the Fermi energy ($E_F$). We see that $\eta(P_{sp}^{max})$ goes to almost $0.15 (k_B \Delta T)^2/h$. The efficiency at maximum spin power as derived from Eq.~\ref{etapmaxsp}, depends on two factors $ZT|_{sp}$ and $P'$. Since in our case $ZT|_{sp}$ takes quite large values its not surprising that we have a highly efficient spin heat engine in addition to a highly efficient charge based one too. Further, we see that the efficiency $\eta(P_{sp}^{max})$ is maximum $0.15 \eta_c$ for $E_F=30 meV$ but at this Fermi energy the maximum spin power delivered is around $0.07 (k_B \Delta T)^2/h$. However, at Fermi energy close to $35 meV$ the efficiency although slightly lower at $0.1 \eta_c$ the maximum spin power output is $0.1 (k_B \Delta T)^2/h$ . As stated before, we not only need high efficiency but we need to deliver large output spin power too  balancing these two needs implies that operating the spin heat engine at $E_{F}=35 meV$ will satisfy both our needs. {Next, in Fig.~9, we have plotted the heat current $J_Q$ as a function of Fermi energy. In Fig.~9(a), we explain the effect of magnetic impurity on heat current via examining three cases. First case is in absence of magnetic impurity, i.e., the exchange coupling strength term $J=0$. For the second case $J=$finite, but the spin flip probability of the magnetic impurity $F=0$ ($S=m=5/2$), i.e., it acts as a non-magnetic impurity. Finally, for the third case we study the heat current in presence of a magnetic impurity, i.e., $J\neq 0, F\neq 0$ ($S=5/2,m=-5/2,F=\sqrt{5}$). We find that introducing a magnetic impurity does not change the heat current much as shown in Fig.~9(a). However, increasing strain reduces the heat current, 
see Fig.~9(b). Thus, we see that spin polarization is not so important for the heat current.} 

 \begin{figure}[h]
 \includegraphics[width=0.9\textwidth]{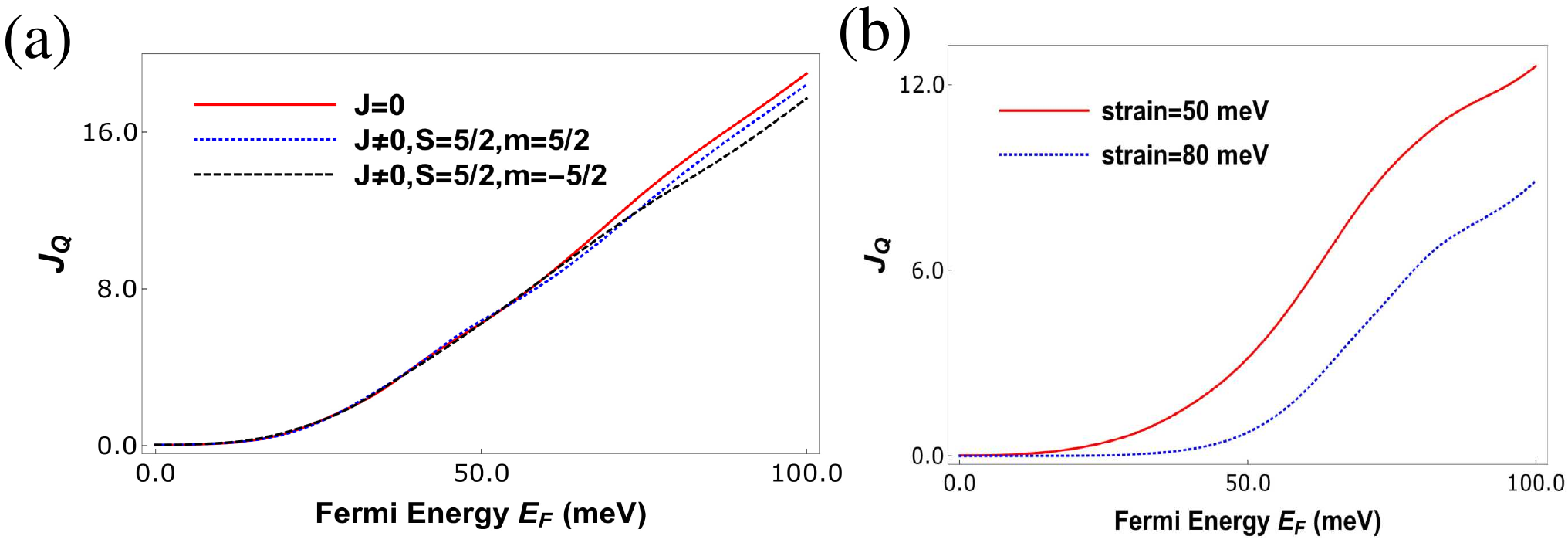}
       \caption{(a) Heat current vs Fermi energy, for three cases- in absence of any impurity, in presence of non-magnetic impurity and in presence of magnetic impurity. The parameters are exchange coupling $J=200$ meV-nm, strain$=30$ meV, $L=60$ nm, (b) heat currents vs Fermi energy in presence of a magnetic impurity, for different values of strains with exchange coupling $J=800$ meV-nm, $L=60$ nm, $S=5/2, m=-5/2$.}
       \label{fig10}
\end{figure}

\subsection{Pure spin current}
\begin{figure}[h]
     {\includegraphics[width=0.6\textwidth]{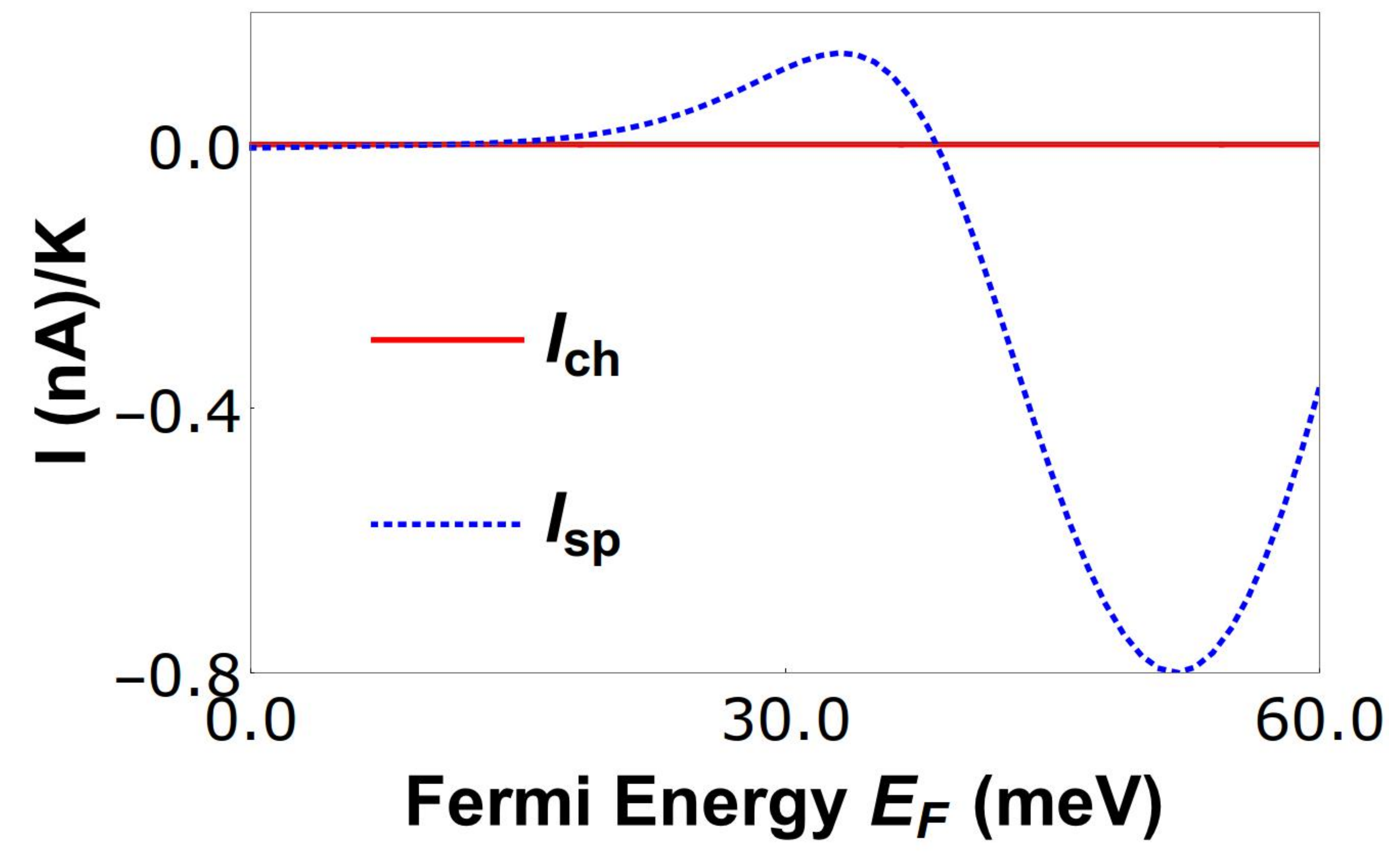}}
       \caption{Charge $I_{ch}$ and spin $I_{sp}$ currents are shown for different parameters than that which works for quantum charge/spin heat engine setups. The parameters are-$L=40$ nm, $T=30$ K, $J=232$ meV-nm, and strain $t=50$ meV.}
       \label{purespinfig}
\end{figure}

In our graphene QSHE, in presence of bias voltage and temperature difference, a pure spin current can be generated by optimizing the parameters. If we set the voltage bias $\mathcal{E}$ and the temperature difference $\Delta T$ such that, the electrical current due to voltage bias and the electrical current generated from temperature difference are exactly equal and opposite to each other then the total charge current will be zero. Though, a finite spin current will be present within the sample, this is the pure spin current. When the total charge current $J_{ch}=0$ in Eq.~(9), we get $\mathcal{E}=-S_{ch}\Delta T$. Substituting this again in the Eq.~(9), we get the pure spin current $J_{sp}=G_{ch}S_{ch}\Delta T(P-P')$. If the polarization $P$ of the spin conductance is different from the polarization $P'$ of the product of $G_{ch}$ and $S_{ch}$, then there will be a pure spin current in our device.  In Fig.~\ref{purespinfig}, we see that the charge current generated is zero, while the spin current is finite. 

\subsection{Graphene bandstructure in presence of strain and spin-flip scattering}
\begin{figure}[h]  
\includegraphics[width=.99\textwidth]{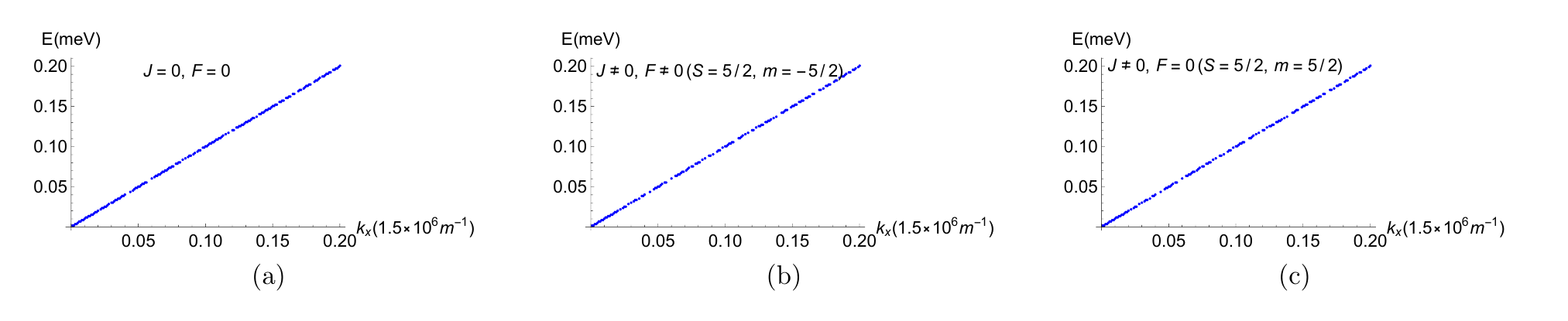}
\caption{ Bandstructure of our model system as function of $k_{x}$ (in $meter^{-1}$) in absence of strain ($t=0$) with length of strain region $L=130\mu m$.}
\label{ff1}	
\end{figure}
\begin{figure}[h]  
\includegraphics[width=.99\textwidth]{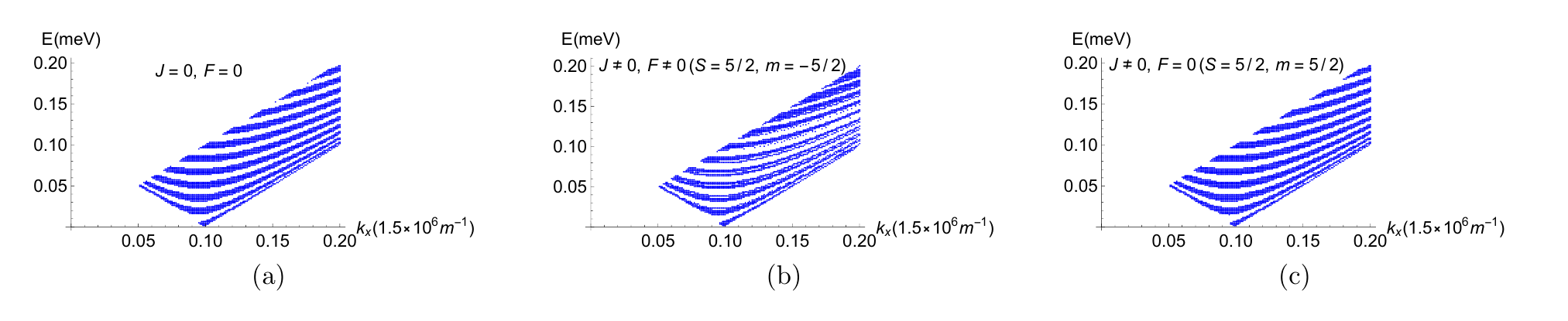}
\caption{ Bandstructure of our model system as function of $k_{x}$ (in $meter^{-1}$) in presence of strain ($t=0.1meV$) with length of strain region $L=130\mu m$.}
\label{ff2}
\end{figure}
To better understand our results we plot the energy bands for 3 cases: (1) in absence of magnetic impurity ($J=0$), (2) in presence of magnetic impurity but without spin flip ($J\neq0$, $F=0$) and (3) in presence of both magnetic impurity and spin flip ($J\neq0$, $F\neq0$) for both unstrained and strained graphene in Figs.~\ref{ff1}, \ref{ff2} respectively. We calculate band structure by considering the wave functions for bound state where the incoming wave in the normal region i.e., $x<0$ region and $x>L$ region do not contribute. The wave function for A sublattice in normal region for K-valley is mentioned in Eq.~25. When we put bound state condition ($\phi\rightarrow i\varphi$) in Eq.~25 we get-\\
For $x<0$-
\begin{equation}
\Psi^{1}_{A}(x)=r_{\uparrow}e^{-i\kappa x}\begin{pmatrix}
                                           1\\
                                           0
                                          \end{pmatrix}\chi_{m}+r_{\downarrow}e^{-i\kappa x}\begin{pmatrix}
                                          0\\
                                          1
                                         \end{pmatrix}\chi_{m+1}
                                         \label{EE}
                                         \end{equation}
where $\kappa=\frac{E}{\hbar v_{F}}\cosh\varphi$. Thus, after putting bound state condition we still get propagating state in Eq.~\ref{EE}. Therefore, we rotate the x-y plane and write the wave function in y-x plane similar to reference\cite{Vitor}. Eq.~25 in y-x plane becomes-\\
For $y<0$-
\begin{equation}
\Psi^{1}_{A}(y)=r_{\uparrow}e^{-i k_{y} y}\begin{pmatrix}
                                           1\\
                                           0
                                          \end{pmatrix}\chi_{m}+r_{\downarrow}e^{-ik_{y} y}\begin{pmatrix}
                                          0\\
                                          1
                                         \end{pmatrix}\chi_{m+1}
                                         \label{EE1}
                                         \end{equation}
where $k_{y}=\frac{E}{\hbar v_{F}}\sin\phi$. Similarly, we can write the wave function for other regions ($0<y<L$ and $y>L$) in y-x plane.
After putting bound state conditions, the bound state wave functions for A and B-sublattice in each region (normal and strained) for K-valley can be written as-\\
For $y<0$-
\begin{equation}
\Psi^{1}_{A}(y)=r_{\uparrow}e^{\kappa y}\begin{pmatrix}
                                  1\\
                                  0
                                 \end{pmatrix}\chi_{m}+r_{\downarrow}e^{\kappa y}\begin{pmatrix}
                                  0\\
                                  1
                                 \end{pmatrix}\chi_{m+1}
\label{en1}
                                 \end{equation}
\begin{equation}
\Psi^{1}_{B}(y)=r_{\uparrow}e^{\kappa y}\begin{pmatrix}
                                      e^{\varphi}\\
                                      0
                                     \end{pmatrix}\chi_{m}+r_{\downarrow}e^{\kappa y}\begin{pmatrix}
                                     0\\
                                     e^{\varphi}
                                     \end{pmatrix}\chi_{m+1}
\end{equation}
in region $0<y<L$-
\begin{equation}
\Psi^{2}_{A}(y)=a_{\uparrow}e^{ik_{y}y}\begin{pmatrix}
                                        1\\
                                        0
                                       \end{pmatrix}\chi_{m}+b_{\uparrow}e^{-ik_{y}y}\begin{pmatrix}
                                       1\\
                                       0
\end{pmatrix}\chi_{m}+a_{\downarrow}e^{ik_{y}y}\begin{pmatrix}
                                        0\\
                                        1
                                       \end{pmatrix}\chi_{m+1}+b_{\downarrow}e^{-ik_{y}y}\begin{pmatrix}
                                                                               0\\
                                                                               1
                                                                              \end{pmatrix}\chi_{m+1}
\end{equation}
\begin{equation}
\Psi^{2}_{B}(y)=a_{\uparrow}e^{ik_{y}y}\begin{pmatrix}
                                        e^{i\theta}\\
                                        0
                                       \end{pmatrix}\chi_{m}+b_{\uparrow}e^{-ik_{y}y}\begin{pmatrix}
                                       e^{-i\theta}\\
                                       0
\end{pmatrix}\chi_{m}+a_{\downarrow}e^{ik_{y}y}\begin{pmatrix}
                                       0\\
                                       e^{i\theta}
                                       \end{pmatrix}\chi_{m+1}+b_{\downarrow}e^{-ik_{y}y}\begin{pmatrix}
0\\
e^{-i\theta}
\end{pmatrix}\chi_{m+1}                                     
\end{equation}
and for $y>L$
\begin{equation}
\Psi^{3}_{A}(y)=t_{\uparrow}e^{-\kappa y}\begin{pmatrix}
                                          1\\
                                          0
                                         \end{pmatrix}\chi_{m}+t_{\downarrow}e^{-\kappa y}\begin{pmatrix}
                                         0\\
                                         1
                                         \end{pmatrix}\chi_{m+1}
                                         \end{equation}
\begin{equation}
\Psi^{3}_{B}(y)=t_{\uparrow}e^{-\kappa y}\begin{pmatrix}
                                          e^{-\varphi}\\
                                          0
                                         \end{pmatrix}\chi_{m}+t_{\downarrow}e^{-\kappa y}\begin{pmatrix}
                                         0\\
                                         e^{-\varphi}
\end{pmatrix}\chi_{m+1}
\label{en2}
\end{equation}
From the bound state wave equations it follows that $(\frac{E}{\hbar v_{F}})^2=k_{x}^2-\kappa^2=(k_{x}-t)^2+k_{y}^2$, $k_{x}=\frac{E}{\hbar v_{F}}\cosh\varphi$, $\kappa=\frac{E}{\hbar v_{F}}\sinh\varphi$, $k_{x}-t=\frac{E}{\hbar v_{F}}\cos\theta$, $k_{y}=\frac{E}{\hbar v_{F}}\sin\theta$. The boundary conditions at $y=0$-
\begin{equation}
i\hbar v_{F}[\Psi^{2}_{B}(y=0)-\Psi^{1}_{B}(y=0)]=\frac{J}{2}\vec{s}.\vec{S}[\Psi^{1}_{A}(y=0)+\Psi^{2}_{A}(y=0)]
\label{en3}
\end{equation}
and
\begin{equation}
i\hbar v_{F}[\Psi^{2}_{A}(y=0)-\Psi^{1}_{A}(y=0)]=\frac{J}{2}\vec{s}.\vec{S}[\Psi^{1}_{B}(y=0)+\Psi^{2}_{B}(y=0)]
\end{equation}
and at $y=L$-
\begin{equation}
\Psi^{2}_{A}(y=L)=\Psi^{3}_{A}(y=L) 
\end{equation}
and
\begin{equation}
\Psi^{2}_{B}(y=L)=\Psi^{3}_{B}(y=L) 
\label{en4}
\end{equation}
After substituting the wave functions (Eqs.~\ref{en1}-\ref{en2}) in Eqs. (\ref{en3}-\ref{en4}), at $y=0$ we get-
\begin{align}
{}&(e^{i\theta}+i\alpha m)a_{\uparrow}+(e^{-i\theta}+i\alpha m)b_{\uparrow}+(-e^{\varphi}+i\alpha m)r_{\uparrow}+i\alpha Fr_{\downarrow}+i\alpha Fa_{\downarrow}+i\alpha Fb_{\downarrow}=0\label{en5}\\
{}&i\alpha Fa_{\uparrow}+(e^{i\theta}-i\alpha(m+1))a_{\downarrow}+i\alpha Fr_{\uparrow}-(e^{\varphi}+i\alpha(m+1))r_{\downarrow}+i\alpha Fb_{\uparrow}-(i\alpha(m+1)-e^{-i\theta})b_{\downarrow}=0\\
{}&(1+i\alpha me^{i\theta})a_{\uparrow}+(1+i\alpha me^{-i\theta})b_{\uparrow}-(1-i\alpha me^{\varphi})r_{\uparrow}+i\alpha Fr_{\downarrow}e^{\varphi}+i\alpha Fe^{i\theta}a_{\downarrow}+i\alpha Fb_{\downarrow}e^{-i\theta}=0\\
{}&(1-i\alpha(m+1)e^{i\theta})a_{\downarrow}+(1-i\alpha(m'+1)e^{-i\theta})b_{\downarrow}-(1+i\alpha(m'+1)e^{\varphi})r_{\downarrow}+i\alpha Fe^{i\theta}a_{\uparrow}+i\alpha Fb_{\uparrow}e^{-i\theta}+i\alpha Fr_{\uparrow}e^{\varphi}=0\nonumber\\
\end{align}
and at $y=L$ we get-
\begin{align}
a_{\uparrow}e^{ik_{y}L}+b_{\uparrow}e^{-ik_{y}L}-t_{\uparrow}e^{-\kappa L}=0\\
a_{\downarrow}e^{ik_{y}L}+b_{\downarrow}e^{-ik_{y}L}-t_{\downarrow}e^{-\kappa L}=0\\
a_{\uparrow}e^{ik_{y}L+i\theta}+b_{\uparrow}e^{-ik_{y}L-i\theta}-t_{\uparrow}e^{-\kappa L-\varphi}=0\\
a_{\downarrow}e^{ik_{y}L+i\theta}+b_{\downarrow}e^{-ik_{y}L-i\theta}-t_{\downarrow}e^{-\kappa L-\varphi}=0\label{en6}\\\nonumber
\end{align}
Eqs.~(\ref{en5})-(\ref{en6}) consists of 8 unknowns: $r_{\uparrow}, r_{\downarrow}, a_{\uparrow}, a_{\downarrow}, b_{\uparrow}, b_{\downarrow}, t_{\uparrow}, t_{\downarrow}$. After eliminating these 8 unknowns from Eqs.~\ref{en5}-\ref{en6}, we will get a one single equation-
\begin{equation}
f(k_{y}L,\theta,\varphi,\alpha,F,m)=0 
\label{en7}
\end{equation}
Since $f(k_{y}L,\theta,\varphi,\alpha,F,m)$ is a large expression, we do not explicitly write it here. The solution of Eq.~\ref{en7} for each value of $k_{x}$ gives the bandstructure of our system. In Fig.~\ref{ff1} we plot the energy bands as function of $k_{x}$ in case of unstrained graphene ($t=0$) for $J=0$, flip and no flip process. We see that in all cases the energy band structure is identical. Similar to Fig.~\ref{ff1} we plot the energy bands as function of $k_{x}$ in case of strained graphene ($t=0.1meV$) in Fig.~\ref{ff2}. We see that band gap opening occurs in the energy bands of strained graphene in contrast to the unstrained graphene. Further, in case of spin flip process there are more number of states in the energy bands as compared to $J=0$ and no spin-flip scattering. The extra number of states help in generating large charge/spin Seebeck coefficient. Since charge and spin thermoelectric properties are proportional to the square of the charge and spin Seebeck coefficient, they increase with increasing Seebeck coefficient. In this way, strain and magnetic impurity help to enhance the performance of our graphene spin heat engine.

{\section{Experimental Realization}}
 Our proposal of a quantum heat engine based on a strained monolayer graphene layer doped with a magnetic impurity is experimentally realizable. There are many theoretical, see Ref.~\cite{castro} which initiated the field of straintronics in graphene and Ref.~\cite{nanoscale} for a recent review, as well as experimental papers, see Refs.~\cite{tmg,chhikara}, which deal with uniaxial strain in monolayer graphene system. There should not be much difficulty in realizing strain in a graphene system. In addition, there are theoretical works which deal with effects of  magnetic impurities on electronic transport in graphene, see Refs.~\cite{firoz,abolfath,maruri}. In Ref.~\cite{maruri}, it is shown that a delta potential approximation of a rectangular barrier magnetic impurity in graphene can be a very effective model of a magnetic quantum dot(a quantum dot with spin). For a range of  incident angles from -$\pi/6$ to $\pi/6$, it is seen that the difference between the transmissions through delta potential magnetic impurity and that through a rectangular barrier magnetic impurity in graphene is quite small. The graphene based system in Ref.~\cite{maruri} is very similar to our set-up, and the problem too is solved similar to ours, only difference being that there is no strain in Ref.~\cite{maruri}. In Refs.~\cite{jaye,chen}, an extended line defect has been studied in a graphene nanostructure experimentally. These line defects can be replaced by a magnetic quantum dot doped with $Mn^+$ ions to realize a magnetic impurity, see Refs.~\cite{maruri, abolfath}. Ref.~\cite{abolfath} is an experimental work which shows how doping $Mn^+$ ions into semiconductor quantum dots realizes magnetic quantum dots. Further, magnetic quantum dots have been experimentally realized in  graphene recently, see Ref.~\cite{npjqm}. Since in the aforesaid papers, people have worked on similar systems, thus the applied aspect of our work is evidently realizable. The amount of strain applied in our system is very small. The maximum strain used in our system is $110 meV$, which is equivalent to $4\%$ strain in graphene. In pristine graphene, maximum $20\%$ strain can be reached without opening a band gap. All the numerical values of different parameters are physically realizable and are used in other works also, see Refs.~\cite{castro, dolphas, firoz}. 
 
In a monolayer graphene sample, a local strain can be introduced by depositing the graphene sample onto a homogeneous substrate with different geometrical patterns drawn on it (like grooves, creases, steps, or wells etc). These geometrical patterns are drawn only on the central region of the substrate, thus, creating a finite strain limited to the central region of the graphene sample. In the other regions, there are no geometrical patterns drawn thus creating no strain. These different geometrical patterns, drawn on the substrate, interact differently with the graphene sheet, generating different strain profiles, see Ref.~\cite{ni}. Strain can also be introduced by stretching, compressing or suspending the central region of graphene layer only without affecting the unstrained region, see Ref.~\cite{balandin}. In Ref.~\cite{balandin}, only the central region of the graphene sample is suspended across a wide trench in a Silion substrate, generating a finite strain which is  limited to the central region alone, see figure in Box.~1(a) (at page 573 of the Ref.~\cite{balandin}). Since regions 1 and region 3 of the graphene sheet are not suspended, strain in these regions will be zero. In this way, one can change the in plane hopping amplitude for the central region of the graphene sample while it remains unchanged for the unstrained regions. In our model, strain is strictly applied to the central region of the graphene sample. However, experimentally it could be possible that the finite strain is not limited to the central region only. It can gradually decrease to zero as one moves from strained  to unstrained region, see Ref.~\cite{dolphas}. In our model, we have finite and constant strain in the central region of graphene sample, while in regions 1 and 3 it is completely zero. This consideration of the sharp drop of strain potential at the boundary between two regions is for our convenience only. We can consider a small slope for strain between the strained and unstrained regions also. However, theoretical calculations predict that the tunneling probability for  electrons will not be that affected whether the strain potential is perfectly sharp or has a small slope at the boundary between strained and unstrained regions. This has also been discussed in other works, see Refs.~\cite{periera,martino}. Moreover, in recent times significant advances have been made in controlling strain in a graphene sample via strain engineering, see Ref.~\cite{ni}.
 {
\begin{center}
\begin{table}
\caption{Comparison of graphene spin heat engine with quantum spin Hall heat engine}
\begin{tabular}{ |c|c|c|}
 \hline
&Helical heat engine, Ref.~\cite{hel}&This work\\
\hline
$P^{max}_{ch}$ &$0.8 \frac{(K_B\Delta T)^2}{h}$&$0.16 \frac{(K_B\Delta T)^2}{h}$\\
\hline
$P^{max}_{sp}$ &$10 \frac{(K_B\Delta T)^2}{h}$&$0.1 \frac{(K_B\Delta T)^2}{h}$\\
\hline
$\eta(P_{ch}^{max})$ & $0.28 \eta_c$&$0.48 \eta_c$\\
\hline
$\eta(P_{sp}^{max})$&$0.4\eta_c$&$0.1\eta_c$\\
\hline
\end{tabular}\\
Note-Helical heat engine relies on edge modes while graphene spin heat engine relies on ballistic modes. 
\end{table}
\end{center}} 
\section{Conclusion and Perspective on ballistic verses edge modes based quantum spin heat engines}
{We have shown in this manuscript that a strained graphene layer, embedded with a magnetic impurity, can act both as a charge as well as a spin heat engine, with higher efficiency than other similar systems.} In Table 1, we compare our graphene QSHE with quantum spin Hall based heat engine\cite{hel}. We see that the maximum charge power for our graphene QSHE is $0.16(k_B \Delta T )^2 /h$, which is less than that of the quantum spin Hall heat engine (maximum charge power of $0.8(k_B \Delta T)^2 /h$). However, efficiency at the maximum charge power in graphene QSHE is $0.48\eta_c$, larger than that observed in quantum spin Hall heat engine, which is $0.28\eta_{c}$. Further, maximum spin power for graphene QSHE is $0.1(k_B \Delta T)^2 /h$ while efficiency at that spin power is $0.1\eta_c$ which are less
than that of the quantum spin Hall heat engine. Thus from Table 1 we see that in some respects ballistic modes are better and in some respects edge modes are better for thermoelectric applications. 

\section{Acknowledgments}
This work was supported by funds from Science \& Engineering research Board, New Delhi, Govt. of India, Grant No. EMR/20l5/001836.  CB  acknowledges Aspen Center for Physics (supported by National Science Foundation grant PHY-1607611 and also a Simons Foundation grant) for support wherein part of this work was completed.
   
 \section*{Author contributions statement}
C.B. conceived the proposal,  A.M. did the calculations of thermoelectric properties on the advice of C.B.,  A.M., S.P and C.B. analyzed the results and wrote the paper.  A.M, S.P and C.B. reviewed the manuscript.
\section*{Competing interests statement}
The authors have no competing  interests.
\end{document}